\documentstyle{cargese}\input epsfig

\let\footnote\savefootnote
\let\footnotetext\savefootnotetext 
\let\footnoterule\savefootnoterule 
\normallatexbib

\catcode`\@=11
\def\mref#1{\ifx\und@fined#1
{need to supply reference \string#1.}
\else #1 \fi}
\catcode`\@=12 %
\let\lref=\def

\newcommand\cA{{\cal A}} 
\newcommand\cC{{\cal C}} \newcommand\cD{{\cal D}}
\newcommand\cE{{\cal E}}
 \newcommand\cG{{\cal G}}
\newcommand\cH{{\cal H}} 
 
\newcommand\cL{{\cal L}} \newcommand\cM{{\cal M}}

\newcommand\inbar{\vrule height1.5ex width.4pt depth0pt}
\newcommand\IC{\relax\,\hbox{$\inbar\kern-.3em{\rm C}$}}
\newcommand\IQ{\relax\,\hbox{$\inbar\kern-.3em{\rm Q}$}}
\newcommand\IR{\relax{\rm I\kern-.18em R}}
\newcommand\IF{\relax{\rm I\kern-.18em F}}
\newcommand\IP{\relax{\rm I\kern-.18em P}}
\newcommand\IH{\relax{\rm I\kern-.18em H}}
\def\ZZ{\relax{\sf Z\kern-.4em Z}}
\newcommand{\beq}{\begin{equation}}
\newcommand{\eeq}{\end{equation}}
\newcommand{\eel}[1]{\label{#1}\end{equation}}
\newcommand{\bea}{\begin{eqnarray}}
\newcommand{\eea}{\end{eqnarray}}
\newcommand{\eeal}[1]{\label{#1}\end{eqnarray}}
\newcommand{\beac}{\begin{equation}\begin{array}{rcl}}
\newcommand{\eeacn}[1]{\end{array}\label{#1}\end{equation}}
\newcommand{\non}{\nonumber}

\newcommand{\eq}[1]{(\ref{#1})}
\newcommand{\del}{\partial}

\newcommand\Coeff[2]{{#1\over #2}}
\newcommand\coeff[2]{\relax{\textstyle {#1 \over #2}}\displaystyle}

\newcommand\attac[1]{\Bigl\vert
{\phantom{X}\atop{{\rm\scriptstyle #1}}\phantom{X}}}

\newcommand\figinsert[4]
{\begin{figure}[htb]
\newcommand\figsize{#3}
\epsfysize\figsize
\centerline{\epsffile{#4}}
\vskip -.2cm
\caption{\leftskip 1pc \rightskip 1pc
\baselineskip=10pt plus 2pt minus 0pt {\sl {#2}}}
\label{fig:#1}
\end{figure}}
\newcommand\taueff{\Delta_{{\rm eff}}}

\newcommand\Tr{{\rm Tr}}
\newcommand\tr{{\rm tr}}
\newcommand\LCS{\cL^{\rm (CS)}}
\newcommand\Aroof{{\widehat{\cA}}}
\newcommand\Lroof{{\widehat{\cL}}}
\newcommand\taus{\tau_{IIB}}
\def\Cp#1{C^{(#1)}}
\def\Bp#1{B^{(#1)}}
\def\Li{{\cal L}i}
\def\l{\lambda}

\begin{document}


\articletitle[Tests of Heterotic/F-Theory Duality]{
On the Heterotic/F-Theory\\ \ Duality in Eight Dimensions}

\chaptitlerunninghead{On the Heterotic/F-Theory Duality}

\author{W.\ Lerche}


\affil{CERN, Geneva, Switzerland}         
%
%

\begin{abstract}
We review quantitative tests on the duality between the heterotic
string on $T^2$ and $F$-theory on $K3$. On the heterotic side, certain
threshold corrections to the effective action can be exactly computed
at one-loop order, and the issue is to reproduce these from geometric
quantities pertaining to the $K3$ surface. In doing so we learn
about certain non-perturbative interactions of 7-branes.
\end{abstract}


\section{Introduction}

One of the most basic dualities in string theory is the one between the
heterotic string, compactified on the two-torus $T^2$, and $F$-theory
on $K3$ \cite{Fth}; indeed other dualities can be derived from
this duality in $d=8$.  In fact, the higher the uncompactified
space-time dimension is, the simpler the structure of non-perturbative
effects becomes, and related to that, the less complicated the
structure of the moduli space is. As we will see, in $d=8$ the
structure is simple enough for exactly computing certain
non-perturbative quantities, but still complicated enough to obtain
functionally non-trivial results.

The point is that certain pieces of the effective action can be
computed {\it exactly} at one-loop order in the heterotic formulation.
This is the reason why this model provides an ideal framework for
studying non-trivial brane interactions; in the past, it has been very
successfully applied to study brane interactions in type I and matrix
strings (see eg., \cite{BFKOV,ko,typeI}).  Our aim, on the other hand,
is to show how these coupling functions can be reproduced in
$F$-theory, ie., from \index{K3 geometry}$K3$ geometry.  This has also
a direct interpretation in terms of certain Type IIB $D$-brane
interactions, which gives another motivation for studying this subject.

Usually, when studying interactions between $D$-branes, one considers
idealized situations where one focuses one a single pair of (possibly
stacks of) branes. In addition, these branes are usually mutually
``local'', i.e., they can be simultaneously described at weak coupling,
and so treated with methods of conformal field theory.

However, for making quantitative tests of string dualities involving
highly non-trivial functions, the full {\it global} structure of
$D$-brane interactions becomes important as well -- that is, the
influence of all the other branes that are usually considered as far
away. Some of the branes are necessarily non-local with
respect to each other, which precludes any conformal field theoretic
treatment of the full theory.

Therefore, in order to exactly determine the functional dependence of
the relevant interactions, new non-perturbative methods are called for.
Guided by the exact results that can be obtained in the heterotic
formulation, we will show how these interactions can be described in
geometrical terms. This approach has been presented in refs.\
\cite{WLSS,LSWA,LSWB}, and this is what we will --partially-- review
from section 3.4 onwards.

However, before we will come to that, we will first review some
simple facts about the structure of the coupling functions
under consideration. Subsequently we will present a brief
introduction to the relevant aspects of $F$-theory.

\section[Exact Heterotic One-Loop Amplitudes]
{BPS Saturated, Exact Heterotic Amplitudes at One-Loop Order}

One certainly cannot expect to compute any given piece of the effective
lagrangian exactly. It is in general just very special couplings,
namely typically those which are anomaly-related and to which only
BPS-states contribute, that are amenable to an exact treatment. In the
present situation with 16 supercharges in eight dimensions, the
canonical BPS-saturated amplitudes \cite{WL,BK} involve four external
gauge bosons (and/or gravitons that we will not consider here).

Supersymmetry relates parity even ($i\xi\,F^n$) and parity odd
(${\theta\over2\pi} F \wedge F\wedge .. F$) sectors, and one can
conveniently combine the theta-angle and the coupling constant $\xi$
into one complex coupling, $\taueff$. In particular, when compactifying
the heterotic string on $T^2$, the effective \index{threshold
couplings}threshold couplings $ \taueff(T,U) \equiv i\xi(T,U)+
\Coeff1{2\pi} \theta(T,U)$ become highly non-trivial functions
depending on the usual torus K\"ahler and complex structure moduli,
$T$ and $U$.

As mentioned before, in the heterotic string picture these couplings
are exact at one-loop order; this is simply because there are no
instantons that could possibly contribute (apart from the world-sheet
instantons whose effects are captured in the one-loop computation). The
couplings are in fact directly related \cite{WL} to the heterotic
\index{elliptic genus}elliptic genus \cite{ellg}, which is given by the
Ramond partition function in the presence of a non-vanishing background
gauge field strength, roughly: $\Aroof_{}(F,q)\sim\Tr_R(-1)^{J_0}
q^{L_0}e^{F}$. More precisely, the couplings  are typically given by
modular integrals of the form:
\beq
{\rm Re}[\taueff(T,U)]\, F\wedge ...F\ \,
\sim\ \int {d^2\tau\over {\tau_2}}
Z_{(2,2)}(T,U,q)\Aroof_{}(F,q) \attac{8-form}\ ,
\eel{ellgen}
where $Z_{(2,2)}$ is the partition function of the two-torus $T^2$:
$$
Z_{2,2}(T,U,q)\ =\ \sum_{p_L,p_R} q^{{1\over2} |p_L|^2}
\bar q^{{1\over2}|p_R|^2}\ ,
$$
with $p_L={1\over \sqrt{2T_2 U_2}}(m_1+m_2\bar U+n_1\bar T+n_2\bar
T\bar U)$, $p_R={1\over \sqrt{2T_2 U_2}}(m_1+m_2\bar U+n_1 T+n_2 T\bar
U)$ (we switched off the Wilson lines here). The evaluation of modular
integrals of type \eq{ellgen}\ is quite an art that has been discussed
at length in \cite{HM,ko,WLSS,LSWA}, but won't be touched upon here.
Essentially, each of such integrals results in a certain
holomorphically factorized, \index{Borcherds product}Borcherds product
type of automorphic function:
\bea
\taueff(T,U)\ &=&\
\ln[\Psi]\ ,\qquad {\rm where}\cr
\Psi &=& (q_T)^a (q_U)^b ~ \prod_{(k,l)>0} \left(1- {q_T}^k {q_U}^l
\right)^{ c(k l)}\ ,
\eeal{borch}
for some $a,b$. Here, $q_T = e^{2 \pi i T}$, $q_U = e^{2 \pi i U}$, the
product runs over $k>0,\ l\in \ZZ\ \ \wedge\ \ k=0, \ l>0$ in the
chamber $T_2\equiv {\rm Im}T>U_2\equiv {\rm Im}U$,  and $c(n)$ are the
expansion coefficients of a certain nearly holomorphic and
(quasi-)modular form, $\cC(q)\equiv\sum c(n) q^n$. The precise form of
this ``counting function'' $\cC(q)$ depends on the specific gauge
couplings that are considered.

Specifically, if we switch off all the Wilson lines so that we
have $E_8\times E_8'$ non-abelian gauge symmetry, we have \cite{ko,LSWA}:
\bea
\Delta_{E_8E_8'}(T,U) &=& - 48\, \ln[\Psi]  \ , \ {\rm with}\ a=-2\ ,
b= 0\ {\rm and\ counting\ function} \cr
&&\qquad\qquad\qquad\cC(q)={1\over 12}{1\over \eta^{24}} \Big[ E_2 E_4
- E_6 \Big]^2(q) \\ \Delta_{E_8E_8}(T,U) &=& - 24\, \ln[\Psi]  \ , \
{\rm with}\ a= 8\ , b= 12\ {\rm and\ counting\ function} \cr
&&\qquad\qquad\qquad\cC(q)={1\over 12} {E_4\over \eta^{24}}\Big[ E_2^2
E_4 - 2 E_2  E_6 + E_4^2 \Big](q) \ ,
\non
\eeal{E8E8}
where $E_n(q)$ are the usual Eisenstein series of the corresponding
modular weight. Moreover, for those couplings for which the field
strengths $F=\{F_T,F_U\}$ are superpartners of the torus moduli, an
extra structure \cite{WLSS,LSWB}\ emerges: namely these couplings
satisfy non-trivial integrability conditions and so can be obtained as
fourth derivatives of the following holomorphic
prepotential:\footnote{The very existence of a holomorphic prepotential
hints at the existence of a yet unknown superspace formulation of the
theory, in which the prepotential would figure as the effective
lagrangian.}
\beq
\cG(T,U)\ \sim\ \sum_{(k,l)>0}
c(kl)\,\Li_5({q_T}^k {q_U}^l)\ ,
\eel{prepot}
with counting function given by $\cC(q)={E_4^2\over\eta^{24}}(q)$.
Above, the polylogarithm is defined by $\Li_a(z)=\sum_{p>0} {z^p \over
p^a}$ $(a \geq 1)$.

An physically interesting feature of these couplings in the $T,U$
subsector is that they have logarithmic singularities, 
consider for example:
\bea
\taueff^{(TTUU)}(T,U)&\equiv& 
(\del_T)^2\,(\del_U)^2\, \cG(T,U)\\
&=&{1\over 2\pi i}\ln[J(T)-J(U)] + {1\over 2\pi i}
\ln[\Psi_0(T,U)]\non ,
\eeal{logsing}
where $\Psi_0(T,U)$ is some cusp form that stays finite over the whole
of the moduli space. Similar to the analogous situation in four
dimensions \cite{dieter,enhancements}, the modular invariant
$J$-functions encode the gauge symmetry enhancements pertaining to the
compactification torus $T^2$: $SU(2)$ for $T=U$, $SU(2)\times SU(2)$ at
$T=U=i$ and $SU(3)$ at $T=U=\rho\equiv e^{2\pi i/3}$, and in particular
reflect the charge multiplicities of the states becoming light near the
singularities. Specifically, near the $SU(2)$ locus the coupling
behaves like:
$$
\taueff^{(TTUU)}(T,U)\ \sim\ \ln[\sqrt{\alpha'}a]\ ,\qquad\
a\equiv \coeff1{\sqrt{\alpha'}}(T-U)\ ,
$$
and similar for the other gauge groups. This is the expected
behavior of the one-loop field theory effective action, with cutoff
scale given by $\alpha'$.

The issue is to reproduce the threshold coupling functions
$\taueff(T,U)$ in the dual $F$-theory compactification on $K3$ (we will
actually consider one-dimensional slices of the moduli space with
constant $U$). For this, we will first briefly review some of
the relevant basic features on $F$-theory that we will need.

\section{Review of $F$-Theory}

\subsection{Elliptic Fibrations}

$F$-theory compactifications \cite{Fth}\  are by definition
compactifications of the type IIB string with non-zero, and in general
non-constant string coupling -- they are thus intrinsically
non-perturbative. $F$-theory may also seen as a construction to
geometrize (and  thereby making manifest) certain features pertaining
to the $S$-duality of the type IIB string.

To explain this in somewhat more detail, let us first recapitulate the
most important massless bosonic fields of the type IIB string. From the
NS--NS sector, we have the graviton $g_{\mu\nu}$, the antisymmetric
2-form field $B$ as well as the dilaton $\phi$; the latter, when
exponentiated, serves as the coupling constant of the theory. Moreover,
from the R--R sector we have the $p$-form tensor fields $\Cp p$ with
$p=0,2,4$. It is also convenient to include the magnetic duals of these
fields, $\Bp6$, $\Cp6$ and $\Cp8$ ($\Cp4$  has self-dual field
strength). It is useful to combine the dilaton with the axion into one
complex field:
\beq
\taus\ \equiv\ \Cp0+ie^{-\phi}\ .
\eel{taudef}
The $S$-duality then acts via projective $SL(2,\ZZ)$ transformations
in the canonical manner: $\taus\to{a\taus+b\over c\taus+d}$ with
$a,b,c,d\in\ZZ$ and $ad-bc=1$. Furthermore, it acts via simple
matrix multiplication on the other fields if these are grouped into
doublets $\left(\Bp2\atop\Cp2\right)$,  $\left(\Bp6\atop\Cp4\right)$
(while $\Cp4$ stays invariant).

The simplest $F$-theory compactifications are the highest dimensional
ones, and simplest of all is the compactification of the type IIB string on
the 2-sphere, $\IP^1$.\footnote{Six and in particular for dimensional
compactifications are much more complicated than the eight
dimensional one discussed here, and any discussion of them would be
beyond the scope of this lecture.} However, as the first Chern class
does not vanish: $c_1(\IP^1)=-2$, this by itself cannot be a good,
supersymmetry preserving background. The remedy is to add extra
7-branes to the theory, which sit at arbitrary points $z_i$ on the
$\IP^1$, and otherwise fill the 7+1 non-compact space-time dimensions.
If this is done in the right way, $c_1(\IP^1)$ is cancelled, thereby
providing a consistent background.

\figinsert{encirc}{
Encircling the location of a 7-brane in the $z$-plane leads to a jump of the perceived type IIB string coupling, $\taus\to\taus+1$.
}{.9in}{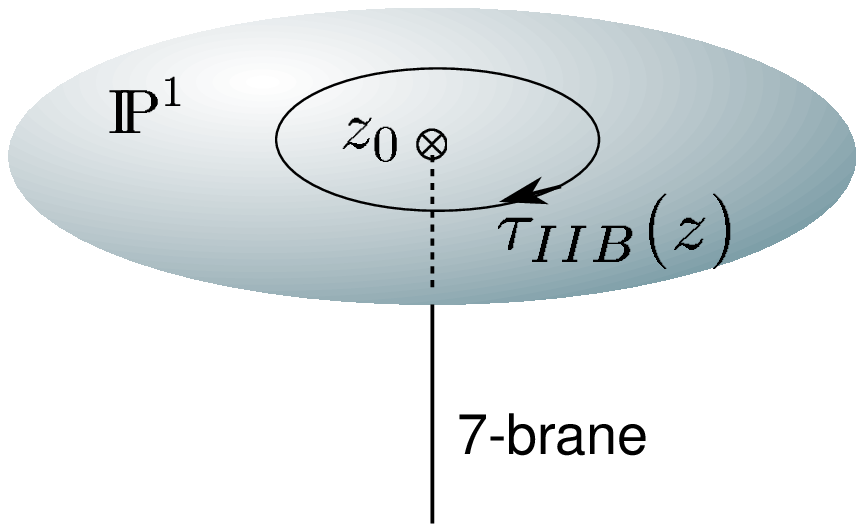}

To explain how this works, consider first a single $D7$-brane located
at an arbitrary given point $z_0$ on the $\IP^1$. A $D7$-brane carries
by definition one unit of $D7$-brane charge, since it is a unit source
of $\Cp8$. This means that is it magnetically charged with respect to
the dual field $\Cp0$, which enters in the complexified type IIB
coupling in \eq{taudef}. As a consequence, encircling the plane
location $z_0$ will induce a non-trivial \index{monodromy}monodromy,
that is, a jump on the coupling -- see Fig.~1.1. But this then implies
that in the neighborhood of the $D7$-brane, we must have a {\it
non-constant} string coupling of the form: $\taus(z)={1\over2\pi
i}\ln[z-z_0]$; we thus indeed have a truly non-perturbative situation.

In view of the $SL(2,\ZZ)$ action on the string coupling \eq{taudef},
it is natural to interpret it as a modular parameter of a two-torus,
$T^2$, and this is what then gives a geometrical meaning to the
$S$-duality group \cite{Fth}. Since, as we have seen, this modular
parameter $\taus=\taus(z)$ is not constant over the $\IP^1$
compactification manifold, the shape of the $T^2$ will accordingly vary
along $\IP^1$. The relevant geometrical object will therefore not
be the direct product manifold $T^2\times \IP^1$, but rather a 
\index{fibration}{\it fibration} of $T^2$ over $\IP^1$ -- see Fig.~1.2.

\figinsert{fibration}{
Fibration of an elliptic curve over $\IP^1$, which in total
makes a $K3$ surface. At 24 points the fibers and therefore the
string coupling become singular, and this is where the 7-branes are
located.
}{1.7in}{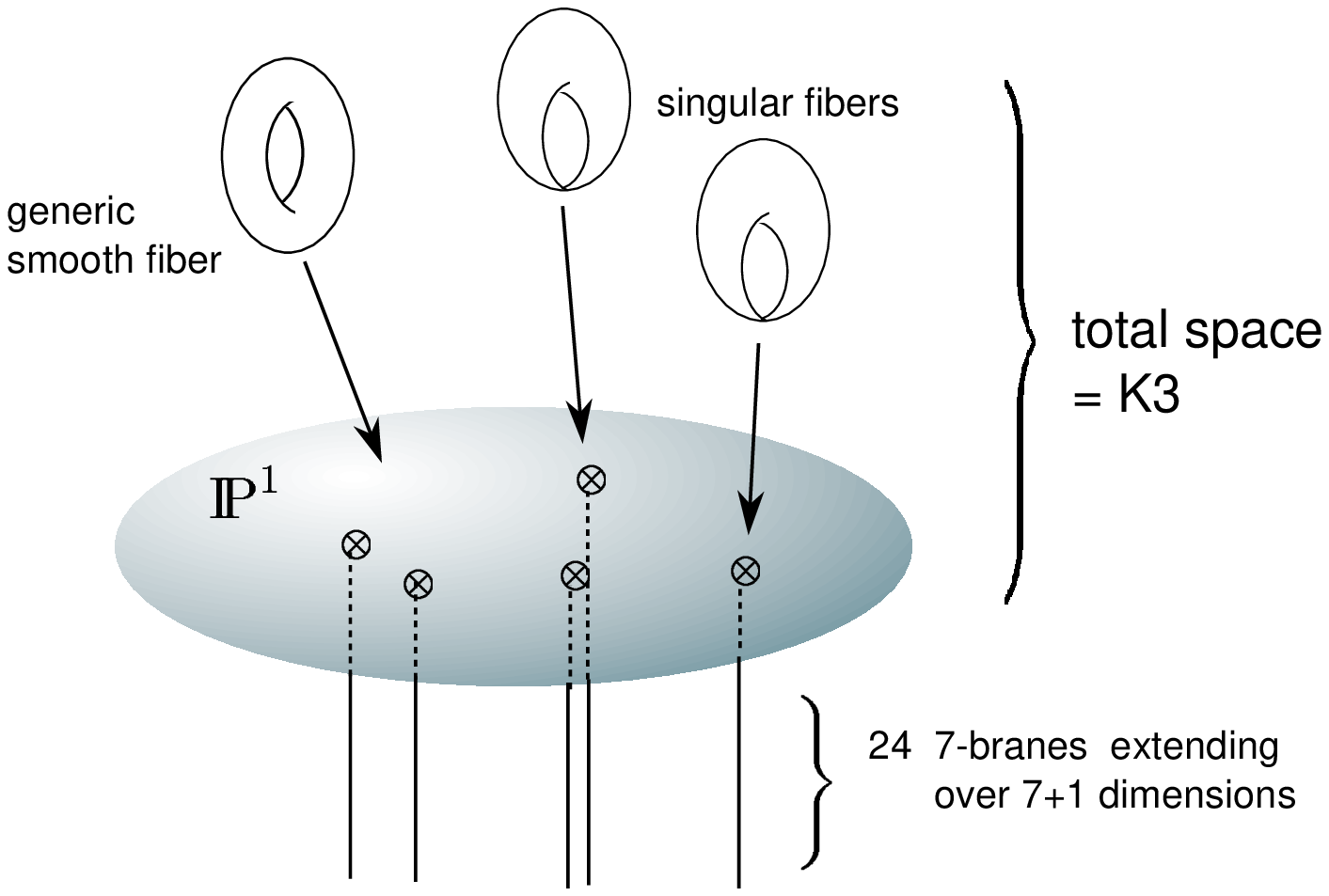}

The logarithmic behavior of $\taus(z)$ in the vicinity of a 7-brane
means that the $T^2$ fiber is singular at the brane location. It is
known from mathematics that each of such singular fibers contributes
$1/12$ to the first Chern class. Therefore we need to put 24 of them in
order to have a consistent type IIB background with $c_1=0$. The
mathematical data: ``$T^2$ fibered over $\IP^1$ with 24 singular
fibers'' is now exactly what characterizes the $K3$
surface;\footnote{More precisely, an ``elliptically fibered'' $K3$;  we
assume here singularities of the simplest canonical type. See ref.\
\cite{PAK3}\ for a physicist's review of the $K3$ manifold.}  indeed
it is the only complex two-dimensional manifold with vanishing first
Chern class (apart from $T^4$).

The $K3$ manifold that arises in this context is so far just a formal
construct, introduced to encode of the behavior of the string coupling
in the presence of 7-branes in an elegant and useful way. One may
speculate about a possible more concrete physical significance, such as
a compactification manifold of  a yet unknown 12 dimensional
``$F$-theory'' \cite{Fth}. The existence of such a theory is still
unclear, but not really important for our purposes; all we need the
$K3$ for is to use its intriguing geometric properties for computing
physical quantities (the quartic gauge threshold couplings, ultimately).

In order to do explicit computations, we first of all need a concrete
representation of the $K3$ surface. Since the families of $K3$'s in
question are \index{elliptic fiber}elliptically fibered,  the natural
starting point is the two-torus $T^2$. It can be represented in the
well-known ``Weierstra\ss'' form:
\beq
W_{T^2}\ =\ y^2+x^3+x f+g\ =\ 0\ ,
\eel{weier}
which in turn is invariantly characterized by the $J$-function:
\beq
J\ =\ {4 (24 f)^3\over 4 f^3+27 g^2}\ .
\eel{Jinvt}
An elliptically fibered $K3$ surface can be made out of \eq{weier}\ by
letting $f\to f_8(z)$ and $g\to g_{12}(z)$ become polynomials  in the
$\IP^1$ coordinate $z$, of the indicated orders. The locations $z_i$ of
the 7-branes, which correspond to the locations of the \index{singular
fiber}singular fibers where $J(\taus(z_i))\to\infty$, are then
precisely where the discriminant
\bea
\Delta(z)&\equiv& 4 {f_8}^3(z)+27 {g_{12}}^2(z)\cr
&=:&\prod_{i=1}^{24}(z-z_i)
\eeal{discr}
vanishes. 

\subsection{Monodromies and Locality}

Note that the polynomials $f_8(z)$ and $g_{12}(z)$ in \eq{weier}\
together have exactly 18 independent free parameters. It is indeed a
mathematical fact that the \index{moduli space}moduli space of elliptic
$K3$ manifolds is 18 complex-dimensional and furthermore locally given
by
\beq
M\ =\ {SO(18,2)\over SO(18)\times SO(2)}\ .
\eel{mspace}
This just happens to be the same as the moduli space of the heterotic
string on $T^2$ (which includes the torus moduli $T,U$ besides the 16
Wilson lines)~! This coincidence of moduli spaces was one of the
primary motivations \cite{Fth}\ for postulating the duality between
$F$-theory on $K3$ (or type IIB on $\IP^1$ with 24 7-branes) and the
heterotic string on $T^2$.

However, \eq{discr}\ tells us that there are indeed 24 plane locations,
and because we have only 18 independent parameters, the locations of
the 7-branes cannot be all independent, but must be to some degree
correlated. Moreover, the heterotic compactification has 16+2=18 $U(1)$
factors (besides the graviphotons),  while we have on the $F$-theory
side 24 7-branes, each of which carries locally a $U(1)$ factor.
Therefore the degrees of freedom of the 24 branes must somehow be
restricted globally, though locally all the branes look the same.

The point is that there are in fact many types of 7-branes, labelled by
their electric and magnetic charges $(p,q)$.\footnote{Mathematically,
this label refers to the homology class of the vanishing 1-cycle
$\gamma$ that characterizes the singular elliptic fiber $T^2$. That is,
$\gamma=p\alpha+q\beta$, where $\alpha,\beta$ form a symplectic basis
of $H_1(T^2,\ZZ)$.} A $D7$-brane is by definition a brane on which a
fundamental type IIB string can end, and by convention carries $(p,q)$
charge of $(0,1)$. Conversely, a 7-brane on which a $D1$ string can end
has $(p,q)=(1,0)$, and the generic $(p,q)$ brane is a 7-brane
on which a $(p,q)$ string \cite{JS}\ can end.

The $SL(2,\ZZ)$ S-duality group acts on the charge labels $^t(p,q)$ in
the obvious manner, and all the $SL(2,\ZZ)$ orbits of a given brane
have locally identical properties. Indeed the monodromy induced by
encircling a $(p,q)$ brane is simply the corresponding $SL(2,\ZZ)$
conjugate of the monodromy $\cM_{(0,1)}=\left({1\ 1\atop0\ 1}\right)$
of a single $D7$-brane, and takes the following form:
\beq
\cM_{(p,q)}\ =\ \pmatrix{1+pq & q^2 \cr -p^2 & 1-p q\cr}\ \in\ SL(2,\ZZ).
\eel{Monodr}
Global consistency then requires that the total monodromy
on the $\IP^1$ base must be trivial:
\beq
\prod_{i=1}^{24}\cM_{(p_i,q_i)}\ =\ {\bf1}\ ,
\eel{global}
which obviously forbids all 24 branes to simultaneously be of type $(0,1)$.

A given pair of 7-branes is said to be ``mutually local''
if their monodromies commute, which is when
\beq
p_1q_2-p_2q_1\ =\ k\in\ZZ
\eel{DZW}
vanishes. Then they can be simultaneously treated at weak coupling,
using methods of ordinary conformal field theory. If on the other hand
the intersection number $k\not=0$, then the branes are mutually
''non-local'' or ``dyonic'',  and cannot simultaneously be described at
weak coupling. In particular, the naive addition of the $U(1)$ gauge
groups that locally live on each brane, is then ill-defined.  Since the
global consistency condition \eq{global}\ turns out to admit only a
maximum of 18 mutually commuting monodromies, it is thus clear that we
cannot expect to see more than 18 independent $U(1)$ factors in the
full theory, even though each of the 24 branes carries one of such
factors locally. Therefore there is generically no weak coupling, or
CFT description description that would be valid for the complete system
of 24 branes.

\subsection{Gauge Symmetries and Kodaira Singularities}

We have mentioned in the preceding section that the moduli space $\cM$
\eq{mspace}\ of elliptic $K3$'s is the same as of the heterotic string
compactified on $T^2$; at generic points in $\cM$ the gauge symmetry is
simply $U(1)^{18}$.  However, there is much more structure in the
theory, in particular at specific sub-loci of $\cM$, extra non-abelian
gauge symmetries can appear. This is familiar for the heterotic string,
where for example we can switch off the Wilson lines and thereby
restore the $E_8\times E_8$ gauge symmetry of the ten dimensional
theory. In fact there can appear any combination of ``simply laced''
gauge groups (those which are classified by the labels $A_n$, $D_n$ and
$E_{6,7,8}$), as long as the total rank does not exceed 18.

An immediate question is therefore how such gauge symmetries appear in
the $F$-theory language. For the $A_n\sim SU(n\!+\!1)$ type of groups, the
answer is well-known and simple: if we place $n$ $D7$-planes near each
other, then the open strings that are stretched in all possible ways
between them describe massive charged gauge bosons of $U(n)/U(1)^n$ --
see Fig.~1.3~a). If the branes collide, these $n(n-1)$ strings will have
zero masses and so lead to a gauge enhancement, $U(1)^n\to U(n)$.

\figinsert{gaugenh}{
Enlargement of non-abelian gauge symmetries: a) $U(n)$ is generated by
open strings stretched between $n$ $D7$-branes of type $(0,1)$;  b)~if
we add a pair of $(1,1)$ and $(1,-1)$ branes, then the extra ``indirect''
trajectories extend $U(n)$ to $SO(2n)$; c)~$E_8$ is generated by
colliding $\cE_6$ and $\cH_2$ Kodaira 7-planes.
}{1.0in}{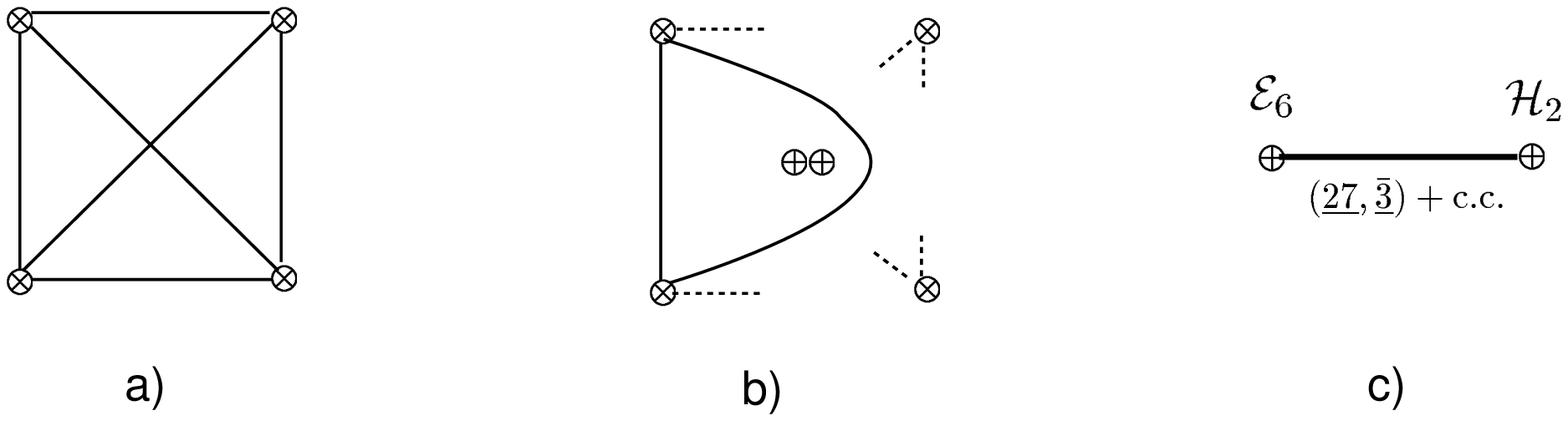}

For the other gauge groups the story is not that simple, however. To
understand it, let us rephrase what we just said in terms of $K3$
geometry. Here each $D7$-brane corresponds to a singular elliptic fiber
with type $(0,1)$ vanishing 1-cycle, located over the same point in the
$z$-plane. Accordingly, the gauge enhancement corresponds in this
language to the collision of \index{singular fiber}singular fibers. In
particular, when all singular fibers are on top of each other, then the
resulting singular fiber will have a worse singularity than we had
before (see Fig.~1.4).

Actually, upon collision not only the elliptic fiber, but also total
space, which is the $K3$ surface, becomes singular or more singular. To
see this, note that the open string stretched between any $D7$-branes
is nothing but a projection of a 2-cycle in the $K3$ - see again
Fig.~1.4.  This 2-cycle can be visualized by dragging the $(0,1)$ cycle
in the fiber along the path of the open string (in other words, the
2-cycle is a fibration of $S^1$ over a line segment).  If the
$D7$-branes collide, then the 2-cycle obviously shrinks to zero size,
and this is precisely what makes the $K3$ singular.

\figinsert{collision}{
One the left we show three singular fibers of type $A_0$. The open
string trajectories are projections of 2-cycles of non-zero volume in
the $K3$. On the right we have collided the singular fibers to form a
singularity of type $A_2$, which is associated with the simultaneous
vanishing of several intersecting 2-cycles in the $K3$.
}{1.9in}{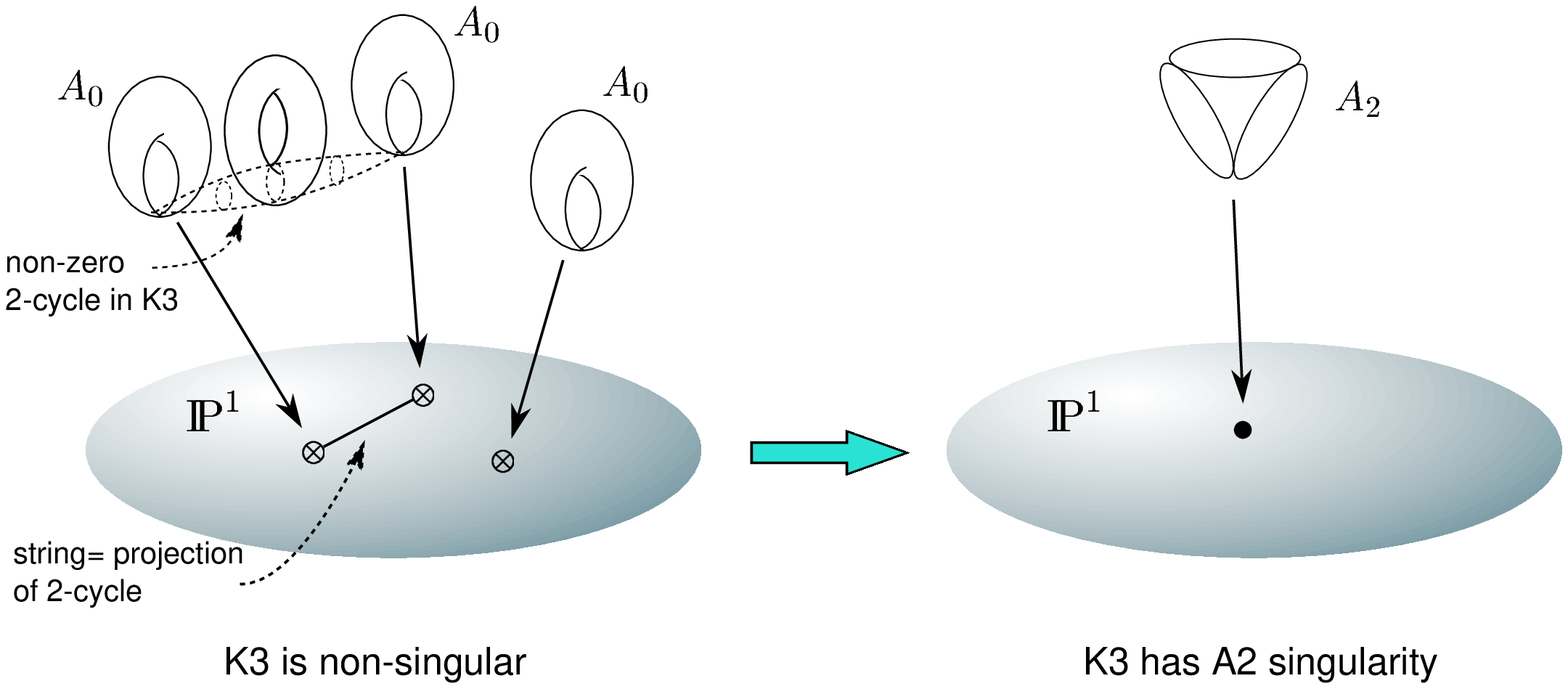}

There is a mathematical \index{Kodaira classification} classification
\cite{Kod}\ of what possible singularities an elliptic $K3$ surface can
have.\footnote{We mean here singularities that can be reached by going
a finite distance in moduli space; there can be other kinds of
singularities like decompactification limits.} These are essentially
given by the well-known \index{ADE singularities}ADE singularities,
however the condition of being elliptic fibrations gives some little
extra structure. The possible elliptic ``Kodaira'' singularities are
listed in Table 1; remember, though, that for a compact $K3$ surface
the total rank that can be achieved is restricted to be less or equal
to~18. The precise manner how the open string trajectories build up the
various charged gauge bosons can in general be quite complicated, and
this is an interesting subject that has been studied in detail in
various papers \cite{junct}; however it is beyond the scope  of this
review.

{\vbox{
{
$$
\vbox{\offinterlineskip\tabskip=0pt
\halign{\strut
$#$\hfil~
&~~$#$~~\hfil
&$#$~~\hfil
&$#$~~\hfil
\cr
\noalign{\hrule}
{\rm Type}
&
{\rm  Gauge\ symmetry}
&
{\rm  7-brane\ content}
&
{\rm Monodromy}
\cr
\noalign{\hrule}
\cA_n
&
U(n+1)
&
A^{n+1}
&
T^{n+1}
\cr
\cD_{n+4}
&
SO(2n+8)
&
A^{n+4}BC
&
S^2T^{-n}
\cr
\cE_{6}
&
E_6
&
A^{5}BC^2
&
(ST)^2
\cr
\cE_{7}
&
E_7
&
A^{6}BC^2
&
S
\cr
\cE_{8}
&
E_8
&
A^{7}BC^2
&
ST
\cr
\cH_{0}
&
U(1)
&
AC
&
(ST)^{-1}
\cr
\cH_{1}
&
U(2)
&
A^2C
&
S^{-1}
\cr
\cH_{2}
&
U(3)
&
(AC)^2
&
(ST)^{-2}
\cr
\noalign{\hrule}}
\hrule}$$
\vskip-10pt
\noindent{\bf Table 1:}
{\sl
Kodaira classification of elliptic singularities ($n\geq0$).
Physically, these correspond to 7-planes which support the respective
$ADE$ gauge symmetries on their world-volumina; many of them cannot be
described in term of weak coupling physics. We also indicated their
(non-unique) decomposition in terms of 7-brane building blocks, where
$A$ denotes a 7-brane of type $(0,1)$, $B$ of type $(1,-1)$ and $C$ of
type $(1,1)$. Moreover we show their $SL(2,\ZZ)$ monodromies in terms
of the usual generators $S$ and $T$.}
}
\vskip10pt}}

Physically, associated to each of these singularities is something what
we may call a ``Kodaira 7-plane'', which may be thought of as
superposition or bound state of ordinary 7-branes. We call these object
``planes'' rather than branes to emphasize that their physical
properties are in general quite different from their ordinary brane
building blocks. For example, as will be important later on, some of
these planes have finite order monodromies, which implies that no
logarithmic branch cuts emanate from them. This in turn means that
these planes have no net $\ZZ$-valued $D$-brane charge, and one may
view their finite $\ZZ_N$ monodromy as a \index{torsion charge}
``torsion'' generalization of ordinary $D$-brane charge. Those planes
may also be regarded as non-perturbative $\ZZ_N$ generalizations of
orientifold planes (which are associated with $\ZZ_2$).

An important physical point is furthermore that for the Kodaira
singularities other than $\cA_n$, the vanishing 1-cycles of the
colliding singular fibers are not all of type $(0,1)$. This means that
all other 7-planes apart from those of type $\cA_n$, involve 7-brane
building blocks of different $(p,q)$ types. In fact one can build all
the possible 7-planes out of just three kinds of building blocks
\cite{junct}, which are indicated in Table 1 as well.

These building blocks are non-local with respect to each other, and
according to our previous considerations, this implies that the
7-planes are generically strongly coupled and cannot be described
by ordinary CFT methods; it is likely that there isn't any lagrangian
description of their world-volumina theories.  The exceptions include
of course the planes of type $\cA_n$, and to some extent also
$\cD_{n+4}$; the latter can be described on a sub-locus of the moduli
space where one puts the $(1,1)$ and $(1,-1)$ branes on top of each
other; the resulting object behaves like an orientifold plane, which can
indeed be dealt with in terms of standard CFT~\cite{Sen}.

\subsection{Constant Coupling Slices of the Moduli Space}

For our purpose of computing effective interactions induced
by the 7-branes, the mutual non-localities of the 24 $(p,q)$ branes
are very inconvenient. However, what we can do to simplify matters
is to restrict to slices of the moduli space where the 24 branes combine into composites that have commuting net monodromies.

Particularly simple are the sub-cases where the monodromies not only
commute, but are also of finite order. These correspond to theories
where the 7-brane charge is cancelled locally, such that the type IIB
string coupling $\taus$ is constant over the $\IP^1$ base.
Remembering from \eq{Jinvt}\ that the coupling is determined by
\beq
J(\taus(z))\ =\ {4 (24 f_8(z))^3\over 4 f_8(z)^3+27 g_{12}(z)^2}
\ \mathop{\equiv}^!\ {\rm const.},
\eel{constTau}
we see that requiring $z$-independence yields three (partially
overlapping) possibilities for splitting up the 24 branes
\cite{Sen,KDSM}:
\vskip .7cm   

\noindent {\bf i)} $g_{12}=0\to J=1728\to\taus=i$: group into eight
$\cH_1$-planes (after fixing three points due to $SL(2,\IC)$
invariance, this gives five independent moduli)

\noindent {\bf ii)} $f_{8}=0\to J=0\to\taus=
\rho\equiv e^{2 \pi i/3}$: group into into twelve
$\cH_0$-planes (nine independent moduli)

\noindent {\bf iii)} $f_{8}={h_4}^2$, $g_{12}=$const.${h_4}^3$
$\to\taus=$arbitrary constant: group into four $\cD_4$-planes (one
independent modulus; this branch intersects branches ii) and iii)).
\vskip .7cm

Upon further specialization, one can have some or all of the $\cH_n$
branes combine into planes with larger gauge symmetries; the possibilities
are summarized in Table 2.

{\vbox{{
$$
\vbox{\offinterlineskip\tabskip=0pt
\halign{\strut
\vrule
{}~$#$
\vrule
& ~$#$~
& ~$#$~
& ~$#$~
& ~$#$~
& ~$#$~
& ~$#$~
& ~$#$~
&\vrule#
\cr
\noalign{\hrule}
{\rm Kodaira\ type}\to
&
\cH_0
&
\cH_1
&
\cH_2
&
\cD_4
&
\cE_6
&
\cE_7
&
\cE_8
&
\cr
\noalign{\hrule}
{\rm Composition}
&
\cH_0
&
\cH_1
&
{\cH_0}^2
&
{\cH_0}^3,{\cH_1}^2
&
{\cH_0}^4
&
{\cH_1}^3
&
{\cH_0}^5
&
\cr
{\rm Torsion\ Charge}
&
\ZZ_6
&
\ZZ_4
&
\ZZ_3
&
\ZZ_2
&
\ZZ_3
&
\ZZ_4
&
\ZZ_6
&
\cr
\taus
&
\rho
&
i
&
\rho
&
{\rm any}
&
\rho
&
i
&
\rho
&
\cr
\noalign{\hrule}}
\hrule}
$$
\vskip-10pt
\noindent{\bf Table 2:}
{\sl
List of 7-planes with finite order monodromies in the $z$-plane, which
do not carry net ($\ZZ$-valued) $D$-brane charge.  They may be viewed
as non-perturbative $\ZZ_N$ generalizations of orientifold planes. We
also list their composition in terms of basic $\cH_k$ building blocks,
as well as the associated constant type IIB string coupling, $\taus$.
}
\vskip10pt}}}

\goodbreak
What we will do, in order to further simplify the problem, is to
consider just one-dimensional slices of these moduli spaces, obtained
by grouping the $\cH_k$ planes into four planes in total. As we will
see, the corresponding geometries will be then pretty easy to deal
with. Explicitly, we will consider from now on the singular $K3$
surfaces defined by the equations $W(x,y,z;\l)=0$, where:
\bea
&&({E_8}^2{H_0}^2):\ \ W=\ y^2 + x^3 + z^5(z-1)(z-\l) \cr
&&({E_7}^2{H_1}^2):\ \ W=\ y^2 + x^3 + x z^3(z-1)(z-\l) \\
&&({E_6}^2{H_2}^2):\ \ W=\ y^2 + x^3 + z^4(z-1)^2(z-\l)^2 \cr
&&({D_4}^2{D_4}^2):\ \ W=\ y^2 + x^3 + z^3(z-1)^3(z-\l)^3\ . \non
\eeal{Kthrees}
The first one yields $E_8\times E_8'$ gauge symmetry, and thus will
correspond to the heterotic model with switched-off Wilson lines of
section 2. The other cases correspond to similar models with certain
Wilson lines switched on; we include them here because we can uniformly
treat all these models in the same way.

\figinsert{Hzero}{
7-plane configuration of the first three models in (1.15),
which describes $K3$ surfaces with elliptic $(\cE_{8-k}\times \cH_k)^2$
singularities. We have indicated the monodromies given by
$\omega=e^{2\pi i/N}$, and also exhibited multiplets of (mutually
non-local) open strings that run between the $\cH_k$ planes. For
$\l(\tau)\rightarrow 1$ the $\cH_k$ planes merge into a single plane,
the strings between them then giving rise to massless charged 
gauge fields that enhance the non-abelian gauge group.
}{1.0in}{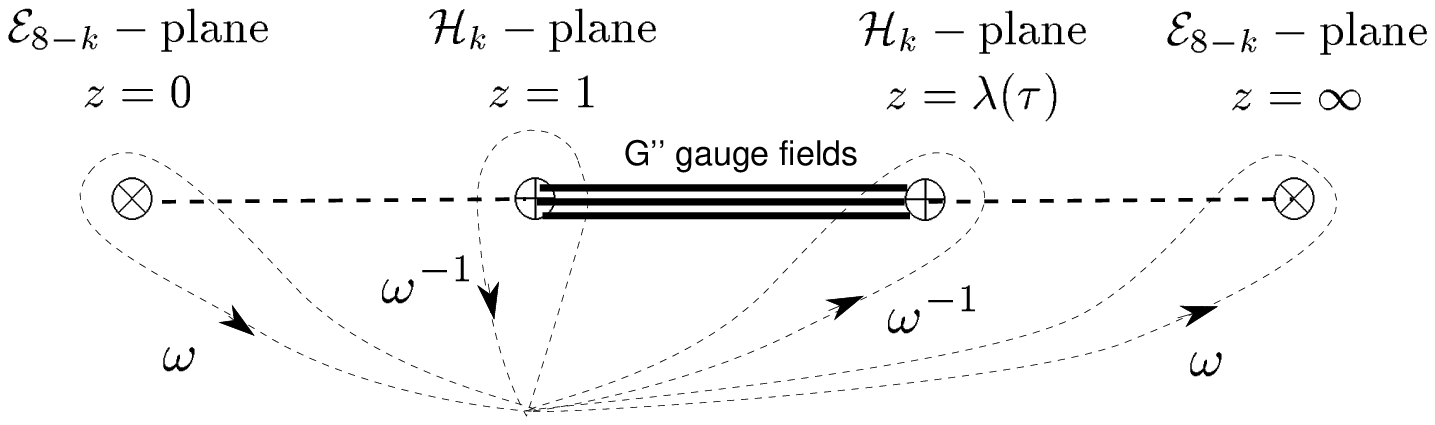}

Each of the surfaces in \eq{Kthrees}\ has two pairs of singular fibers
of the indicated types over the $z$--plane. As convention we have
chosen to put  7--planes of types $\cE_8,\cE_7,\cE_6,\cD_4$ at
$z=0,\infty$, and planes of types $\cH_0,\cH_1,\cH_2,\cD_4$ at
$z=1,\l$, respectively (see Fig.~1.5). Note that the Kodaira
singularity types of these two sets are ``dual'' to each other, in that
the monodromies of the $\cE_{8-k}$ planes and of the $\cH_k$ planes are
inverses of each other; they belong to $\ZZ_N$, $N=2,3,4,6$,
respectively.

In the one-dimensional sub-moduli spaces, two interesting things can
happen. First, a ($\cE_{8-k}$)-- and a $\cH_k$--plane can collide, to
yield an ``$\bar \cE_8$'' singularity of the local form
$y^2+x^3+z^6=0$. As we will see, this corresponds to the
decompactification limit on the heterotic side. Secondly, two
$\cH_k$--planes can collide to produce a 7--plane associated with some
extra non-abelian gauge symmetry $G''$, and precisely which one can be
inferred from Table 2. In other words, the generic non-abelian gauge
symmetry is $(E_{8-k}\times A_k)^2$, which can be enhanced to
$(E_{8-k})^2\times G''$, for $G''=A_2,D_4,E_6$, respectively (for
colliding $\cD_4$ planes there is no further gauge enhancement, as this
also corresponds to the decompactification limit).

\section{Geometric Determination of the Threshold Couplings}

\subsection{Chern-Simons Couplings on Kodaira 7-Planes}

The issue is to compute the functions $\taueff$ \eq{ellgen} via 7-brane
interactions.  Effective interactions in $8d$ space-time are generated
by superimposing world-volume actions, and also by integrating out
exchanges between the 7--branes. While in general very complicated, the
interactions are in the present context reasonably tractable because of
their special anomaly related, parity-odd structure. They arise from
the \index{Chern-Simons term}Chern-Simons terms on the world-volumina
of the 7-branes, via the exchange of $RR$ antisymmetric tensor fields
$\Cp p$.

For a single $D$-brane with $(p,q)=(0,1)$, the relevant tree level
couplings look \cite{GHM}  (for trivial normal bundle):
\beq
\LCS_{D7}\ =\ C\wedge e^{-2iF}\wedge \sqrt{\Aroof(R)} \attac{8-form}\ ,
\eel{Lcs}
where $C\equiv\oplus_{k=0}^4 \Cp {2k}$ is the formal sum over all $RR$
forms, and $\Aroof(R)$ is the Dirac genus.  The couplings for general
$(p,q)$ branes can be obtained by applying $SL(2,\ZZ)$ transformations
on the fields in \eq{Lcs}.

However, due to the generic mutual non-locality of the 24 $(p,q)$
7-branes that we simultaneously have in the theory, it is a priori not
clear how to add up these terms and how to determine what effective
interactions they induce. But as discussed above, we can simply
restrict to sub-moduli spaces where the 24 branes combine into 7-planes
 that are all mutually local, ie., have commuting monodromies. Then all
the contributions can simply be added up.

In order to do so, we will first need to know what the
relevant couplings on the world-volumina on the various kinds of
7-planes are, in analogy to the couplings on a single $D7$-brane
\eq{Lcs}. Because of the mutual locality, the anomalous couplings can
in fact be very easily determined. Specifically, recall  that a
$\cD_4$-plane can be viewed as being composed out of four $D7$-branes
plus one orientifold plane, which are all mutually local. Since a
direct CFT computation gives $\LCS_{O7}\ =\ -4C\wedge
\sqrt{\Lroof(R)}|_{\rm{8-form}}$ \cite{DJM,MSM,OHirz}, where
$\Lroof(R)$ is the Hirzebruch genus, we thus have:
\bea
\LCS_{\cD_4}\ &=&\
C\wedge\left[\tr\big(e^{-2iF}\big)\wedge\sqrt{\Aroof(R)}-
4\sqrt{\Lroof(R)} \right]\attac{8-form}\cr
&=& \Cp4\!\wedge\Big(\coeff12 R^2-2\tr F^2\Big)
\\&&\quad +
\Cp0\!\wedge \Big(\coeff23\tr F^4-\coeff1{12}\tr F^2 \tr R^2
+\coeff1{192}(\tr R^2)^2+\coeff1{48}\tr R^4 \Big)\non
\eeal{D4CS}
Summing over all four  world-volumina indeed exactly reproduces the
(eight dimensional remainder of the) Green-Schwarz term of the
heterotic string, ${\cal L}^{(GS)}=\Cp6\!\!\wedge2(R^2\!-\!\tr
{F_{SO(32)}}^2)\!+\!\Cp2\!\wedge\! X_8({F_{SO(32)},R})$.

The same logic must be valid for the $\cH_0$ and $\cH_1$-planes and
their composites. Even though these planes are associated with strong
coupling and may not have a well-defined lagrangian description of
their world-volume theories, the WZ coupling terms are
topological and independent of the coupling,  and must make sense at
least for anomaly cancelling reasons. Therefore, we can conclude for
the basic building blocks:
\beq
\LCS_{{\cH_0}} = {1\over3} \LCS_{\cD_4}\ ,\qquad  
\LCS_{{\cH_1}} = {1\over2} \LCS_{\cD_4}\ ,
\eel{HnCS}
where the gauge field traces follow implicitly from
the decomposition $SO(8)\to U(1)$ or $U(2)$, respectively.

\subsection{Geometric Interactions on $\ZZ_N$ Curves}

What we are interested in are the non-trivial interactions between the
planes, which should ultimately reproduce the coupling functions
$\taueff(T,U={\rm const})$ of section 2.  The primary perturbative
contributions will arise from massless $\Cp p$ tensor field exchange
between individual planes. The effective interaction will thus depend
on the distances between the various 7-planes in the $z$-plane.

More specifically, the closed string exchange that contributes to the
maximal number of wedge products of field strengths is in the odd $RR$
sector, and is proportional to the Green's function $\Delta$ of a
scalar field on the $z$-plane:
\beq 
\Big\langle\Cp
p_{m_1\dots m_p}(z_1),\Cp{8-p}_{n_1\dots n_{8-p}}(z_2)
\Big\rangle_{{RR^-}}\sim\ \epsilon^{m_1\dots m_pn_1\dots
n_{8-p}}\Delta(z_1,z_2)\ ,
\eel{Ccorrel}
where, as  $z_1\to z_2$:
\beq
\Delta(z_1,z_2)=\ln(z_1-z_2)+{\rm finite}\ .
\eel{Green'sf}
However, in order to obtain functionally exact results, we need to know
the full Green's functions that probe the global structure of the
$z$-plane, and not just their leading singular behavior. This is in
general a complicated problem, but in our setup, where we consider only
planes with finite order monodromies, there is a natural geometric
answer \cite{WLSS,LSWA}. 

\goodbreak
\figinsert{cover}{
Lift of the $z$--plane to a covering Riemann surface. Shown is here the
situation with two $\cE_6$ and two $\cH_2$ planes, which correspond to
$\ZZ_3$ twist fields and anti-twist fields, respectively, located at
the branch points of a genus two curve $\Sigma_2$. We also show an open
string trajectory that contributes to the coupling $\tr
{F_{SU(3)}}^2\tr {F_{SU(3)'}}^2$ (transforming as $(3,\bar 3)$ under
$SU(3)\!\times\!SU(3)'\!\!\subset\! E_6$) and which corresponds to a
$1/3$--period on $\Sigma_2$.}{1.9in}{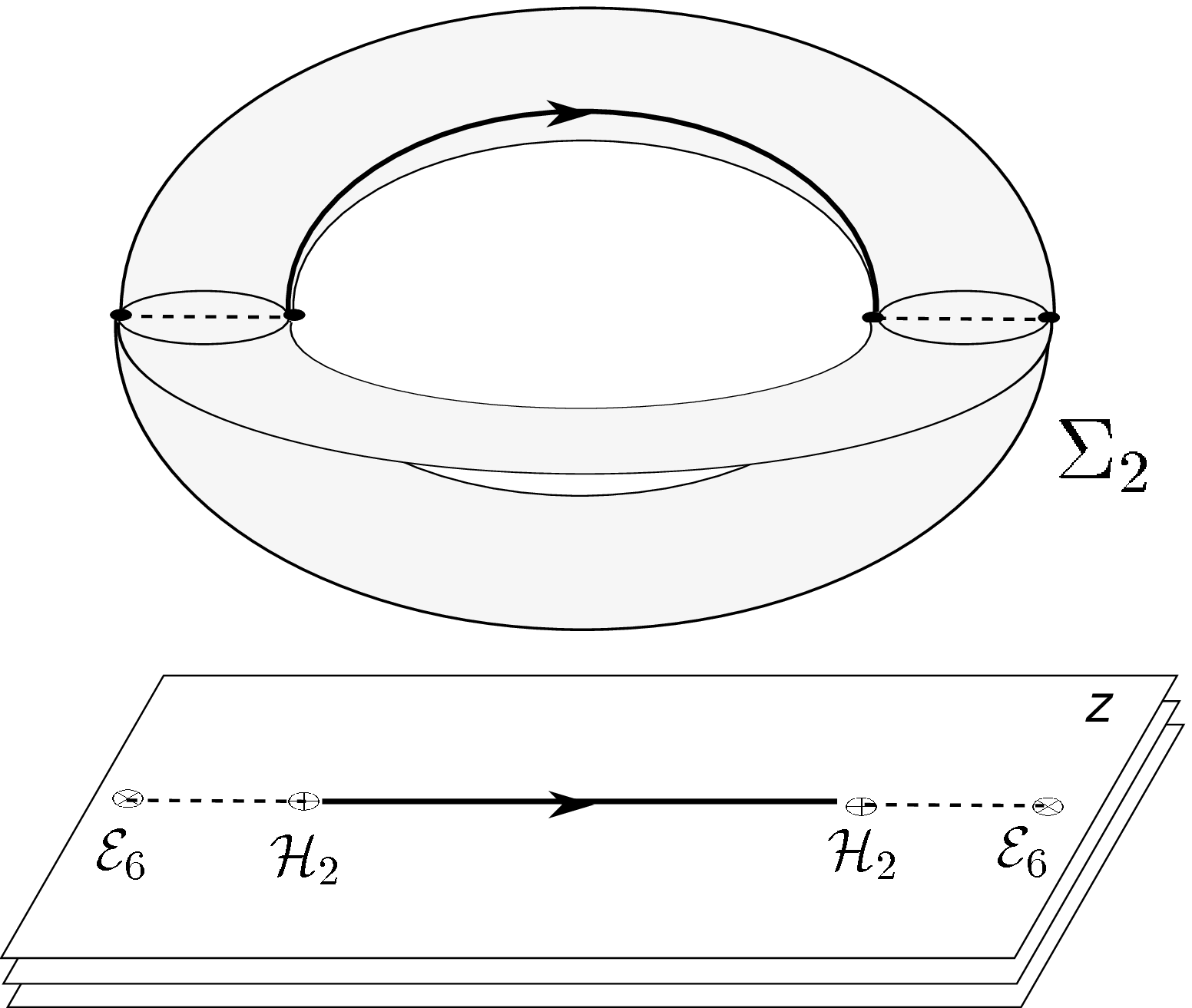}

By definition, monodromies of finite order means that the geometry of
the singular $K3$'s can be described by a finite covering of the
$z$-plane and so effectively reduces to the one of \index{ZN
curves}Riemann surfaces. The four $7$-planes then correspond to the
branch points of these curves (cf. Fig.~1.6). More specifically, for
the four models in \eq{Kthrees}\ one finds the following
$\ZZ_N$-symmetric curves
\beq
\Sigma_{N-1}\,:\qquad x^N \ =\ z^{-1}(z-1)(z-\l)
\eel{zncurves}
of genus $g=N-1$, where $N=6,4,3,2$, respectively.   The requisite
\index{Green's function}Green's functions should therefore simply be
given by appropriate scalar Green's functions on these covering spaces
\cite{ZN}.

The canonical Green's function of a scalar field on a Riemann surface
is known to be given by the logarithm of the ``prime form''
\beq
\Delta^\Sigma_{\rm{prime\atop form}}(z_1,z_2)\ =\ \ln\Big|{
\theta_\delta[\int_{z_1}^{z_2}\vec w\vert\Omega]
\over
\sqrt{\xi(z_1)}\sqrt{\xi(z_2)}
}\Big| -\pi\big[{\rm Im}\int_{z_1}^{z_2}\vec w]\!\cdot\!({\rm
Im}\Omega)^{-1}
\!\!\cdot\! \big[{\rm Im}\int_{z_1}^{z_2}\vec w]\ ,
\eel{primeform}
where $\sqrt{\xi(z)}\equiv\sqrt{{\del\over\del z_i}(\theta_\delta[\vec
z\vert \Omega])\cdot w^i(z)}$ is a some 1/2-differential whose purpose
is to cancel spurious zeros of the numerator theta-function (where
$\delta$ denotes an arbitrary odd characteristic). Indeed, the only
singularity of the prime form is at coincident points, ie.,
$\Delta^\Sigma(z_1\to z_2)\sim\ln[{z_2-z_1\over\sqrt{dz_1}
\sqrt{dz_2}}]+$ finite terms.  By construction, the finite terms
implement the requisite global properties of the Green's functions.

Due to the high degree of symmetry of our $\ZZ_N$ curves
$\Sigma_{N-1}$, much of the information  in \eq{primeform}\ is in fact
redundant for our examples. By explicit computation one can show that
the prime form Green's function, when evaluated between any two of the
branch points, can always be written in the following generic form:
 \beq
\Delta^{\Sigma_{N-1}}_{\rm{prime\atop form}}(z_1,z_2)\ =\
\ln\big[\l^{\alpha_1}(1-\l)^{\alpha_2}(\l')^{\a_3}\big]\ ,
\eel{hauptgreen}
where $z_{1,2}\in
\{0,1,\l,\infty\}$, $z_1\not= z_2$ and the numerical coefficients
$\alpha_i$ depend on the particular choice of $z_1$ and $z_2$.

However, it turns out \cite{LSWA} that these canonical Green's
functions on $\Sigma_{N-1}$ to not capture the full story. They capture
only the exchange of $C$ fields, but miss certain additional instanton
contributions. Namely, loops of $(p,q)$ strings in the $z$-plane will
be closed in general only on the covering surface $\Sigma_{N-1}$, so
that such strings effectively wrap the Riemann surfaces. Wrapping
entire world-sheets of such strings will thus in general generate extra
instanton contributions

These extra contributions can be viewed as modifications of the
canonical Green's functions \eq{hauptgreen}\ into ``effective'' Green's
functions. Indeed, a Green's function is in general ambiguous up to the
addition of a finite piece, and it is this ambiguous piece to which we
can formally attribute those extra non-singular, non-perturbative
corrections. Denoting the extra piece by $\delta$, we can thus write
the threshold coupling functions generically as follows:
\beq
\taueff(\l)
\ =\ \Delta^{\Sigma_{N-1}}_{\rm{prime\atop form}}(\l)+\delta(\l)
\eel{fullGreens}
We will describe further below how to exactly compute the
extra contributions $\delta(\l)$. 

For the time being, note that the above picture applies most directly
to couplings that mix the gauge field strengths of two different
7-planes, which means that they have  the form $\langle
\Cp4\!,\Cp4\rangle$ $\times\tr{F_G}^2 \wedge\tr {F_{G'}}^2$. However,
as one can check on the heterotic side, there are moduli-dependent
corrections also to other eight-form terms in the effective action,
eg., to $(\tr {F_G}^2)^2$, which pertain to a single gauge group factor
living on a {\it single} brane.

\figinsert{global}{
Interactions probing the $\ZZ_3$ torsion piece of $D$-brane charge. A
string junction is shown that contributes to $({\tr F_{SU(3)}}^2)^2$,
transforming as a singlet under $SU(3)\!\times\! SU(3)'\!\subset\!
E_6$. We also show how it lifts to a cycle on the covering curve. The
junction gives rise to a logarithmic singularity when the planes
collide, even though it does not seem to couple locally to the right
$\cH_2$ plane.
}{2.0in}{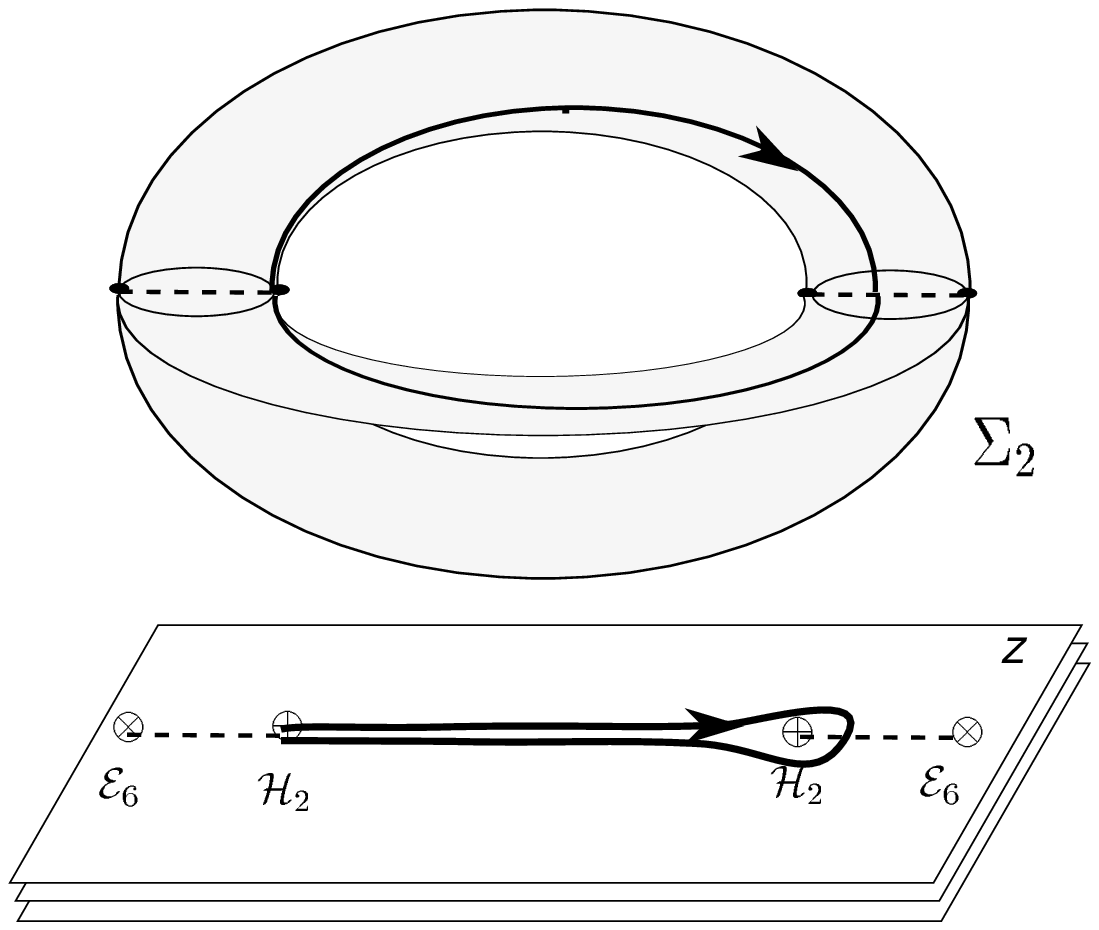}

 In the usually considered situation with CFT description, where one
focuses on pairs of $D$- or orientifold-branes \cite{DJM,MSM,OHirz},
such terms arise from integrating out $\Cp0\!-\!\Cp8$ exchange between
the two branes, each equipped with couplings like $\LCS=Q_7\cdot
\Cp8+\dots+\Cp0\wedge Y_8(F)$, where $Y_8(F)$ is some 8-form
polynomial. This obviously induces a location-dependent correction to
the quartic gauge field coupling of the form  $Q_7\langle\Cp0\!,
\Cp8\rangle \tr Y_8(F_G)$.

However, in the present context, the 7-brane charge is cancelled
locally on every plane so that $Q_7\equiv0$; indeed there is no $\Cp8$
term in \eq{D4CS}. This means that naive $\Cp0\!-\!\Cp8$ exchange
cannot contribute to these couplings.  But how do these (possibly
singular) moduli-dependent corrections, which we explicitly see on the
heterotic side, then arise~? More specifically, how can a given
brane that carries the field strength $F_G$ ``feel'' the presence of
the other brane, while no string ends on that other brane ?

The point is that despite our 7-planes do not have net $\ZZ$-valued
7-brane charge (no logarithmic monodromy), there is still a remnant
left, which is reflected by the finite order $\ZZ_N$ monodromies. It is
this \index{torsion charge} ``torsion'' piece of the $D$-brane charge
that must be responsible for the requisite long-range interactions
\cite{WLSSKod}. This can be seen by analyzing the interactions in terms
of string junctions \cite{junct}. Similar to what is familiar from
orientifold planes,  what one finds are string trajectories that loop
around other planes, rather than coupling to them via the local
Chern-Simons terms in \eq{D4CS}; this is exemplified in Fig.~1.7 for
the $E_6$ model. It thus seems natural to view these interactions  as
analogous to those of ``Alice strings'' \cite{alice}, which do have
long-range interactions but no locally defined charge density.

At any rate, what we learn by studying these couplings is that there
can be non-trivial interactions between 7-branes that should be
attributed the global properties of the multi-valued $z$-plane (which
is best represented by the curves $\Sigma_{N-1}$), rather than solely
to local WZ couplings.

\section{Solution via the Mirror Map}

\subsection[]{Flat coordinates and Moduli Spaces}

The singular $K3$ surfaces in \eq{Kthrees}\ all depend on one geometric
modulus, $\l$. On the other hand, the heterotic models we consider
depend on two moduli $T,U$ (besides the Wilson lines that we keep
frozen and so neglect). In order to compare the geometrical
interactions, which depend on the parameter $\l$ in \eq{hauptgreen},
with the heterotic one-loop results, we therefore --first of all-- need
to know what the map between these moduli is.

The point is that the heterotic moduli are the canonical moduli of a
conformal field theory.  From general reasoning \cite{DVV,CV}\ we know
that the \index{moduli space}moduli space has a flat structure and that
the canonical CFT moduli are the corresponding 
\index{flat coordinates}flat coordinates. 
Therefore, we need to determine what the
flat coordinate $\tau$ is that is associated with the geometric modulus
$\l$. A general way of constructing the flat coordinate is to write
$\tau(\l)$ in terms of the ratio of certain \index{period integrals}
period integrals (which are themselves solutions of certain
linear differential equations, as we will discuss below).

Specifically, the relevant periods pertaining to the  singular $K3$
surfaces $W(x,y,z;\l)=0$ in \eq{Kthrees}\ are obtained by integrating
the unique holomorphic two-form $\Omega^{(2,0)}_{K3}$,
\beq
\varpi_i\ =\ \int_{\gamma_i}\Omega^{(2,0)}_{K3}\ 
\equiv
\int_{\gamma_i}{dx dz\over \del_y W(x,y,z;\l)}
\eel{K3periods}
over a suitable integral basis of 2-cycles $\gamma_i$. We have seen
before that these singular $K3$'s are closely related  to Riemann
surfaces $\Sigma_{N-1}$, and indeed by redefining variables: $x = v
z^{2 (1 - 1/N)}(z-1)^{2/ N} (z - \l)^{2 /N}$, the period integrals
\eq{K3periods}\ factorize into $\int  {d v \over \sqrt{v^3 + 1}}\int
{dz\over z^{1-1/N}(z-1)^{1/N}(z-{\l})^{1/N}}$. The integral over $v$
being a constant normalization that we neglect, the $K3$ period
integrals thus turn into period integrals (now over 1-cycles)
pertaining to the $\ZZ_N$ curves \eq{zncurves}:
\beq
\varpi_i\ =\ \int
{dz\over z^{1-1/N}(z-1)^{1/N}(z-{\l})^{1/N}}\ \equiv
\int_{\gamma_i}\Omega^{(1,0)}_{\Sigma_{N-1}}
\eel{periodint}
These integrals are of canonical hypergeometric type and so
given by linear combinations of
\bea  
\varpi_0(\l)\ &=&\  \int_0^\l\Omega^{(1,0)}_{\Sigma_{N-1}}\ =\
  (-1)^{-2/N} \pi\, {\rm csc}(\coeff\pi N)\,
       {}_2F_1\big(\coeff1N,\coeff1N,1; {\l} \big) \\
\varpi_1(\l)\ &=& \  \int_0^1\Omega^{(1,0)}_{\Sigma_{N-1}}\ =\
   {{\l}}^{-1/N} (-1)^{-2/N} \pi\, {\rm csc}(\coeff\pi N)\,
{}_2F_1\big(\coeff1N,\coeff1N,1; \coeff1{\l} \big)\ .\non
\eeal{curveper}
The flat coordinate is then given, as usual, by the ratio of the
periods:
\bea
\tau(\l)&=&{\varpi_1(\l)\over\varpi_0(\l)}\\
&=& s(0,0,1\!-\!\coeff2N; \l)\ .
\non
\eeal{triangle}

\figinsert{fundR}{
On the left we see a picture of the moduli space of the heterotic
compactification with $E_8\times E_8$ gauge symmetry, given by a
fundamental region $\IF=\IF_\Gamma\cup (S\cdot\IF_\Gamma)\subset\IH^+$. 
The Hauptmodul $\l(\tau)$ maps
this region to the 7-plane geometry shown in the right part of the
figure. The cusps and orbifold point correspond to the various ways the
mobile $\cH_0$ plane can hit the three other planes.
}{1.9in}{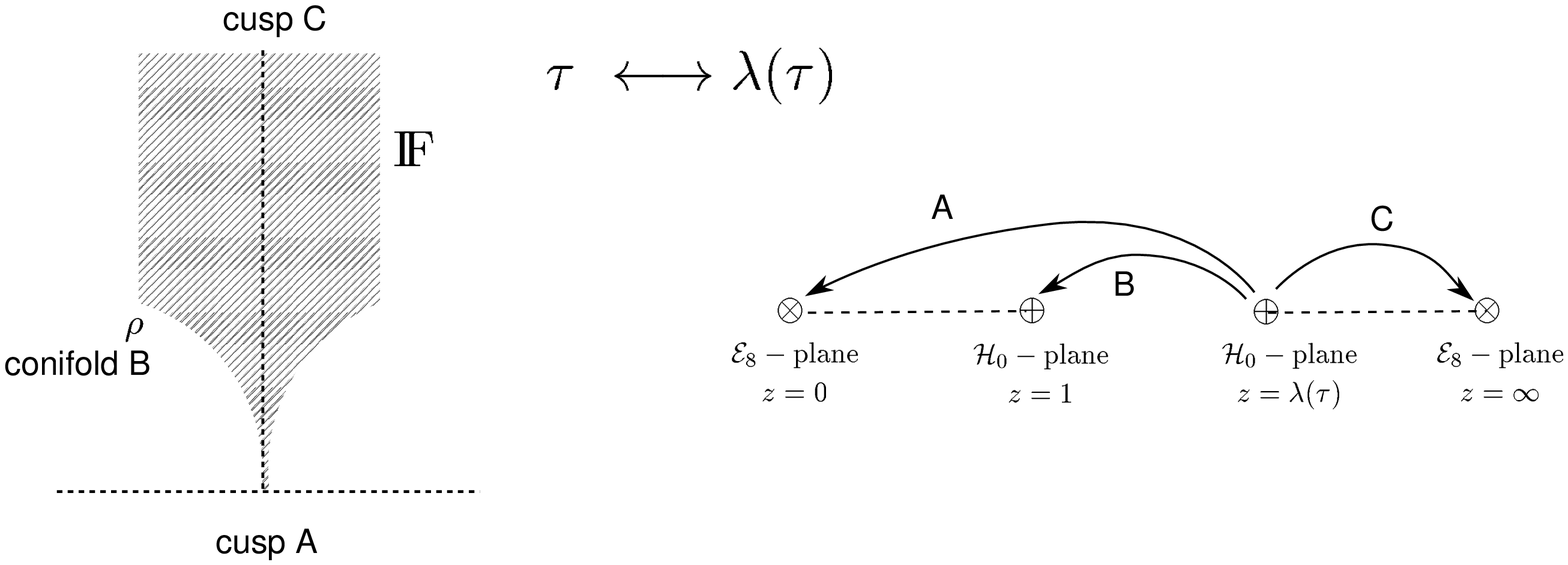}

In \eq{triangle}, $s(a,b,c)$ denotes a triangle function that maps the
complex plane into a \index{fundamental region}fundamental region
$\IF\subset\IH^+$, while its entries $a,b,c$ indicate the angles of
$\IF$. We depict a typical such fundamental region in Fig.~1.8. The two
zero's mean that there are generically two cusps (corresponding to the
decompactification limits $\l\to 0,\infty$), and in addition there is
one orbifold point associated with gauge enhancement, $\l\to 1$
(however, for $N=2$ there are three cusps, which reflects that any two
colliding $\cD_4$ singularities correspond to decompactification).

By the theory of triangle functions, the inverse maps given by
the ``Hauptmoduls'' $\l(\tau)$, can  then be concisely written in terms
of standard modular functions. These, together with the modular
subgroups under which $\l$  is invariant, are listed in Table 3.

{\vbox{
{
$$
\vbox{\offinterlineskip\tabskip=0pt
\halign{\strut
\hfil~$#$
&~~$#$~~\hfil
&$#$~~\hfil
&$#$~~\hfil
\cr
\noalign{\hrule}
N
&
{\rm  mod.\ subgroup}
&
{\rm Hauptmodul\ }\l(\tau=T)
&
\taus=U
\cr
\noalign{\hrule}
6
&
\Gamma_P(2)
&
(\sqrt{-J(\tau)}+\sqrt{1-J(\tau)})^{-2 }
&
\rho
\cr
4
&
\Gamma_0(2)
&
-{1\over{64}}(\coeff{\eta(\tau)}{\eta(2\tau)})^{24} 
&
i
\cr
3
&
\Gamma_0(3)
&
-{1\over{27}}(\coeff{\eta(\tau)}{\eta(3\tau)})^{12}
&
\rho
\cr
2
&\Gamma_0(4)
&
-{1\over{16}}(\coeff{\eta(\tau)}{\eta(4\tau)})^{8}
&
{\rm undet.}
\cr
\noalign{\hrule}}
\hrule}$$
\vskip-10pt
\noindent{\bf Table 3:}
{\sl
Modular subgroups of which $\l(\tau)$ is a modular function for the
$K3$ families in (1.15), and the values of the constant string coupling
$\taus$. They agree with the duality groups of the corresponding dual
heterotic string compactifications. (The modular subgroup
$\Gamma_P(2)\subset SL(2,\ZZ)$ has been discussed in
ref.~\cite{Rankin}).
}
\vskip10pt}}}

If we now identify 
\bea
T&=&\tau\\
U&=&\taus\ \equiv\ {\rm const.}\non
\eeal{identif}
(where the constant type IIB coupling $\taus$ is as listed in Table~3),  
then the above modular subgroups indeed exactly match the
modular subgroups that arise on the heterotic side by switching on the
corresponding Wilson lines.\footnote{The functional map between $T,U$
(with $U$ not being frozen) and the $z$-plane geometry has been
determined for the $E_8\times E_8$ model in refs.\ \cite{CCLM,WLSS}.}

Moreover we can use the map $\l(T)$ to compare the leading
singularities in the geometric Green's function \eq{fullGreens}\  with
the heterotic one-loop results. Note that \eq{hauptgreen}\ carries the
leading logarithmic behavior of the full Green's function when two of
the $\cH_n$ planes collide for $\l\to1$. In particular, for the $E_8$
model we get:
\beq
\taueff(T)\ \mathop{\sim}^{\l\to1}\ \ln[\l(T)]\ \sim \ln[J(T)]
\ ,
\eel{leadingsing}
which captures precisely the singularity of the coupling
$\taueff^{(TTUU)}(T,\rho)$ in \eq{logsing}\ (where we need to set
$J(U)\to J(\rho)\equiv 0$). One can check that similarly for the other
models, there is perfect agreement with the singularities in the
perturbative results on the heterotic side. This yields a first
quantitative, though still superficial test of the duality.

\subsection{String Geodesics and BPS masses}

Remember that the elliptic $K3$ manifold that appears in $F$-theory is
primarily an elegant tool to encode the relevant open string geometry of
non-perturbative type IIB compactifications, and as such should have no
particular physical significance in itself (unless the fictitious twelve
dimensional $F$ theory turns out to exist). Therefore, we should be
able to give the geometrical quantities that we have discussed above a
physical meaning that makes sense more directly in the type IIB
compactification on $\IP^1$.

Indeed the periods and flat coordinates have a direct interpretation in
terms of open strings. Recall \cite{JS}\ that the tension of a $(p,q)$
string (in the canonical metric) is
\beq
T_{p,q}\ =\ {1\over \sqrt{{\rm Im}\taus}}|p+q\taus|\ ,
\eel{tension}
so that the mass of a string stretched along a line $\cC$  is $\int_\cC
T_{p,q}ds$. The line element  has been determined in \cite{GSVY,Sen}
and is given by $ds^2={\rm
Im}\taus|\eta(\taus)|^2\prod_{i=1}^{24}(z-z_i)^{-1/12}dz|^2$. For the
geometries we consider (with four 7-planes and constant type IIB
coupling $\taus$), the mass of a string stretched between any two
planes (at branch points $z_1$ and $z_2$) then simplifies to:
\bea
{m_{p,q}}^2\!\!&=&\!\!\!\!\int_{z_1}^{z_2}\!\!\big\vert
\underbrace{(p+q\taus(z))\eta^2(\taus(z))}_{{\rm const.}}
z^{1/N-1}(z-1)^{-1/N}(z-{\l})^{-1/N}
\!dz\big\vert^2\non\cr
&=&\!|p+q\taus|^2|\eta^2(\taus)|^2\int_{z_1}^{z_2}
\Big|\Omega^{(1,0)}_{\Sigma_{N-1}}\Big|^2\ ,
\eeal{simplint}
which coincides up to a numerical prefactor with the period integral
\eq{periodint}\ (strictly speaking, period integrals are over closed
homology cycles, while here the open strings are stretched along
half-cycles. The difference is just in the normalization). In fact only
ratios of periods have an invariant meaning, whence we have to divide
\eq{simplint}\ by the fundamental period $\varpi_0$. To see that this
provides the correct normalization, consider in the $E_8\times E_8$
model the mass of open strings stretched between the two $\cH_0$
planes. It is proportional to
\bea
{m_{p,q}}^2
&\sim& |p+q\taus|^2\left|
{\int_{1}^{\l}\Omega^{(1,0)}_{\Sigma_{5}}
\over
\int_{0}^{\l}\Omega^{(1,0)}_{\Sigma_{5}}}
\right|^2 \cr
&=& |p+q\rho|^2\left|{\tau-\rho\over \tau-\bar\rho}
\right|^2\ .
\eeal{gaugemass}
With the identification \eq{identif}: $\taus\equiv\rho$, $\tau=T$, this
exactly reproduces the known heterotic mass formula for the
$SU(3)/U(1)^2$ gauge bosons that become massless as the planes collide
(as $\l\to 1$ or $T\to\rho$). The open strings that correspond to these
gauge bosons are sketched in Figs. 1.5 and 1.6; they have $(p,q)$
charges given by $\pm(1,0)$, $\pm(0,1)$ and $\pm(1,-1)$, respectively,
for which \eq{gaugemass}\ implies that they all have the same mass.

One can check that the BPS masses of stretched strings match also in
the other examples those of the corresponding winding and momentum
states in the heterotic string compactifications on $T^2$; a
detailed analysis has been presented in ref.\ \cite{barrozo}.

\subsection[Open-Closed String Mirror Map]
{Mirror Map acting between Open and Closed String Sectors}

Note that in the previous section, we tacitly used a
different language as before: namely we used the concept of
stretched {\it open} strings, while previously we had discussed
$C$-field exchange between 7-planes, which is primarily 
a {\it closed} string concept. Indeed the mass of stretched
open strings is best parametrized by the flat coordinate $T$,
on the other hand closed string interactions between 7-planes
depend on their physical locations in the $z$-plane, and thus
are more naturally parametrized by the geometrical modulus $\l(T)$.

\figinsert{dual}{
Dual interpretations of the same string diagram, obtained by
time-slicing the world-sheet in two orthogonal ways: either in terms of
tree level closed string exchange between the planes, leading to a
contribution $<C,C>\sim\ln[z_1-z_2]$. Or as one-loop diagram involving
stretched open strings, leading to $\sum {Q_i}^4\ln[m_i]$. The
functional relationship between these expressions is essentially
governed by the mirror map, $\tau\leftrightarrow \l(\tau)$.
}{1.3in}{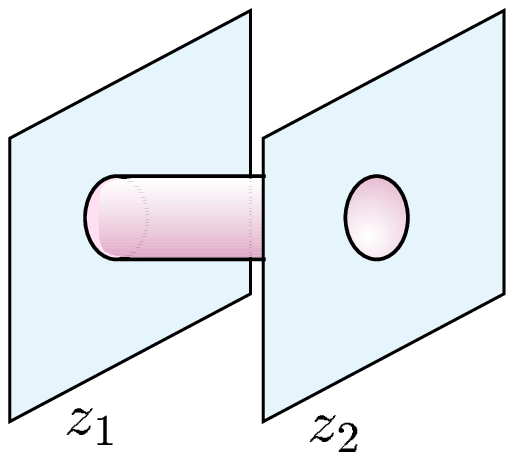}

That a given physical string process can have different interpretations
in terms of open or closed strings is of course a well-known, basic
fact of string perturbation theory -- see Fig.~1.9. To see
this better in the present context, let us recapitulate the origin
of the logarithmic singularity that arises in the effective action when
two $\cH_0$ planes collide in the $E_8\times E_8$ model. 
It can be seen to arise in following two dual ways:

\noindent {\bf i)} It either arises as leading one-loop effect
involving open strings stretched between the planes:
\beq
\taueff\ \sim\ \sum_{{{\rm states}\atop i}}{Q_i}^4\ln[m_i]\ ,
\eel{logmi}
which gives, using \eq{gaugemass}:
\beq
\taueff\ \sim\ 3 \ln\Bigg[{T-\rho\over T-\bar \rho}\Bigg].
\eel{threegauge}
The numerical factor arises due to the 3 sets of charged
$SU(3)/U(1)^2$ gauge bosons which have the same mass.

\noindent {\bf ii)} Alternatively, the
singularity arises --as discussed before-- from massless closed string
$C$-field exchange between the planes, namely as small-distance
singularity of a Green's function:
\beq
\taueff\ \sim\ \langle\Cp4\Cp4\rangle\ \sim\ \ln[z_1-z_2]\ \sim
\ln[1-\l(T)]\ \sim\ \ln[J(T)].
\eel{logJ}
The $J$-function then expands near the singularity as
\beq
J(T\simeq\rho)\ =\ {\rm const.}\ \Bigg({T-\rho\over T-\bar \rho}\Bigg)^3
+\dots,
\eel{Jexpand}
where the power indeed reproduces the same prefactor as in \eq{threegauge}.
Thus the $J$-function (which pertains to the geometry of the $z$-plane)
``knows'' about the open string states of the theory, and effectively
sums up many contributions in a duality invariant way.

The insight that we can abstract from this is that the map
$\tau\leftrightarrow\l$ between flat and geometric coordinates of the
moduli space can be physically interpreted in terms of a map between
the natural open and closed string moduli. From the view point of $K3$
period integrals, the map $\tau\leftrightarrow\l$ can also be viewed as
\index{mirror map} {\it mirror map} \cite{mirror}\ on
$K3$.\footnote{While for higher dimensional Calabi-Yau $d$-folds the
mirror map acts between different manifolds and their moduli spaces,
torus and $K3$ are ``self-mirror'' and accordingly the mirror map acts
within the same moduli space \cite{PADMKthree}.} However, in the
$F$-theory setup the whole of $K3$ is not really physical, and we
should not consider closed two-cycles but rather their projection on
the $z$-plane, which gives open string geodesics. In this sense, the
r\^ole of the $K3$ mirror map in the physical type IIB compactification
on $\IP^1$ is then played by a map between open and closed string sectors.

\subsection[]
{Picard Fuchs Equations, and their Symmetric Square}

We have seen in the preceding sections that a lot can be learned  by
simply focussing on the leading singularities in the moduli space of
the effective theory. However, for the sake of performing really
non-trivial quantitative tests of the heterotic/F-theory duality, we
should try harder in order to reproduce the exact functional form of
the couplings $\taueff(T)$ from $K3$ geometry. The hope is, of course,
to learn something new about how to do exact non-perturbative
computations in $D$-brane physics.

More specifically, the issue is to eventually determine the extra
contributions $\delta(\l)$ to the geometric Green's functions in
\eq{fullGreens}. Having a priori no good clue from first principles how
to do this, the results of the previous section, together with
experience with four dimensional compactifications with $N=2$
supersymmetry \cite{KV,KLM,KKLMV}, suggest that somehow mirror symmetry
should be a useful tool. It was indeed shown in refs.\
\cite{LSWA,LSWB}\ how mirror symmetry can be formally used to do such
computations, although a satisfying physical insight why it
works has not yet been achieved. We will therefore only briefly
sketch the findings of these works, and refer the interested reader to
them for more details.

The starting point is the observation that threshold couplings of
similar structure appear also in four dimensional, $N=2$ supersymmetric
compactifications of type II strings on Calabi-Yau threefolds.  More
precisely, these coupling functions multiply operators of the form
Tr${F_G}^2$ (in contrast to quartic operators in $d=8$), and
can be written in the form
\beq
\taueff^{(4d)}\ \sim\ 
\ln\big[\l^{\alpha_1}(1-\l)^{\alpha_2}(\l')^{\a_3}\big]+\gamma(\l)\ ,
\eel{4dcouplings}
which looks similar to \eq{fullGreens}. By analyzing \cite{LSWA,LSWB}\
the known results of mirror symmetry computations \cite{KLM,KKLMV}\ in
$d=4$, it is found that the ``extra'' term $\gamma(\l)$ in
\eq{4dcouplings}\ appears also in the dilaton flat coordinate. That is,
it is nothing but the remainder of the dilaton in the large dilaton
limit: $S=-\ln[y]+\gamma$, where $y\sim e^{-S}+...$ is a geometric
coordinate of the underlying CY threefold moduli space.

The dilaton $S$ is a period associated with the CY threefold,  and like
all period integrals, it satisfies a system of linear
differential equations. The idea \cite{LSWA,LSWB}\ is thus to first
isolate a differential equation that is satisfied by $\gamma(\l)$ in
$d=4$, and then to see how to generalize it such as to
obtain a differential equation for $\delta(\l)$ in $d=8$. Furthermore,
this differential equation may then be translated back into geometry,
and this then would hopefully give us a clue about what the relevant
quantum geometry is that underlies those quartic gauge couplings in eight
dimensions.

The starting point is, once again, the families of singular $K3$
surfaces in \eq{Kthrees}. As we have seen above, associated with them
are the period integrals \eq{K3periods}, which evaluate to the
hypergeometric functions given in \eq{curveper}. Generally,
\index{period integral}period integrals satisfy the 
\index{Picard-Fuchs equations}``Picard-Fuchs'' 
linear differential equations \cite{HKTY},
and  for our examples \eq{Kthrees}\ these read:  ${\cal L}_N
\cdot\varpi_i(z)=0$, where
\beq
{\cal L}_N(z)  ~=~ 
{\theta_z}^2 - z\,(\theta_z + 
\coeff1{2N})(\theta_z +\coeff1{2}- \coeff1{2N}) \ .
\eel{kthreePFN}
Here we have made for convenience a change of variables: $z\equiv -4
\l/(1-\l)^2$; moreover, $\theta_z\equiv z{\del\over\del z}$.

The four-dimensional theories are obtained by compactifying the  type
II strings on CY threefolds of special type, namely they are {\it
fibrations} \cite{KLM,AL}\ of the $K3$ surfaces \eq{Kthrees}\ over
$\IP^1$. The size of the $\IP^1$ yields then an additional modulus,
whose associated flat coordinate is precisely the dilaton $S$ (in the
dual, heterotic language; from the type II point of view, it is simply
another geometric modulus). The $K3$-fibered threefolds are then
associated with enlarged PF systems of the form:
\bea
{\cal L}_N(z,y)  &=& 
{\theta_z}({\theta_z}-2\theta_y) -
 z\,(\theta_z + \coeff1{2N})(\theta_z +
\coeff1{2}- \coeff1{2N})\cr
{\cal L}_2(y) \ &=&\ {\theta_y}^2-2y\,(2\theta_y+1)\theta_y\ .
\eeal{CYPF}
Since we are interested in the perturbative, one-loop contributions on
the heterotic side (which capture the full story in $d=8$, in contrast
to $d=4$), we need to consider only the weak coupling limit, which
corresponds to the limit of large base space: $y\sim e^{-S}\to0$.
Though we might now be tempted to drop all the $\theta_y\equiv
y\del_y$ terms in the PF system, we better note that the $\theta_y$
term in ${\cal L}_N(z,y)$ can a non-vanishing contribution,
namely in particular when it hits the logarithmic piece of the dilaton
period, $S=-\ln[y]+\gamma$. As a result one finds that the piece
$\gamma$ that we want to compute satisfies in the limit $y\to0$ the
following {\it inhomogenous} differential equation:
\beq
{\cal L}_N\cdot(\gamma\,\varpi_0)(z)\ =\ \varpi_0(z)\ .
\eel{d=4inhom}

We now apply the inverse of this strategy to our eight dimensional
problem. Since we know from the perturbative heterotic calculation of
section 2 what the exact answer for $\delta$ must be (e.g., \eq{E8E8}),
we can work backwards and see what inhomogenous differential
equation the extra contribution $\delta(\l)$ obeys. What we find after
some tedious computations is that it satisfies:
\beq
{\cal L}_N^{\otimes2}\cdot(\delta\,{\varpi_0}^2)(z)\ =\ {\varpi_0}^2(z)\ ,
\eel{d=8inhom}
whose homogenous operator
\beq
{\cal L}_N^{\otimes2}(z)\ =\ 
{\theta_z}^3 - z\,(\theta_z + 1-\coeff1N)(\theta_z +
\coeff12)(\theta_z +  \coeff1N)\ ,
\eel{symL}
is the ``symmetric square'' \cite{LY,symm2}\ of the $K3$ Picard-Fuchs
operator \eq{kthreePFN}. This means that its solution space is given by
the symmetric square of the solution space of ${\cal L}_N(z)$, i.e.,
\beq
{\cal L}_N^{\otimes2}\cdot
({\varpi_0}^2,\varpi_0\varpi_1,{\varpi_1}^2)\ =\ 0.
\eel{symLSol}

Even though the inhomogenous PF equation \eq{d=8inhom}\ concisely captures
the extra corrections in the eight-dimensional threshold terms,
the considerations leading to this equation have been rather formal
and it would be clearly desirable to get a better understanding of what it 
mathematically and physically means.

Note that in the four dimensional situation, the PF operator ${\cal
L}_N(z)$, which figures as homogenous piece in \eq{d=4inhom}, is  by
construction associated with the $K3$ fiber of the threefold. By analogy,
the homogenous piece of equation \eq{d=8inhom}\ should then tell us
something about the geometry that is relevant in the eight dimensional
situation.  Observing that the solution space \eq{symLSol}\ is given by
products of the $K3$ periods, it is clear what the natural
geometrical object is: it must be the symmetric square Sym$^2(K3)=
(K3\times K3)/\ZZ_2$. Being a \index{hyperkahler manifold}hyperk\"ahler 
manifold \cite{hyperk}, its periods (not subject to world-sheet
instanton corrections) indeed enjoy the factorization property
exhibited by \eq{symLSol}.

\figinsert{K3square}{
Formal similarity of the four and eight-dimensional string
compactifications: the underlying quantum geometry that underlies the
quadratic or quartic gauge couplings appears to be given by three- or
five-folds, which are fibrations of $K3$ or its symmetric square,
respectively. The perturbative computations on the heterotic side are
supposdly reproduced by the mirror maps on these manifolds in the limit
where the base $\IP^1$'s are large.
}{1.4in}{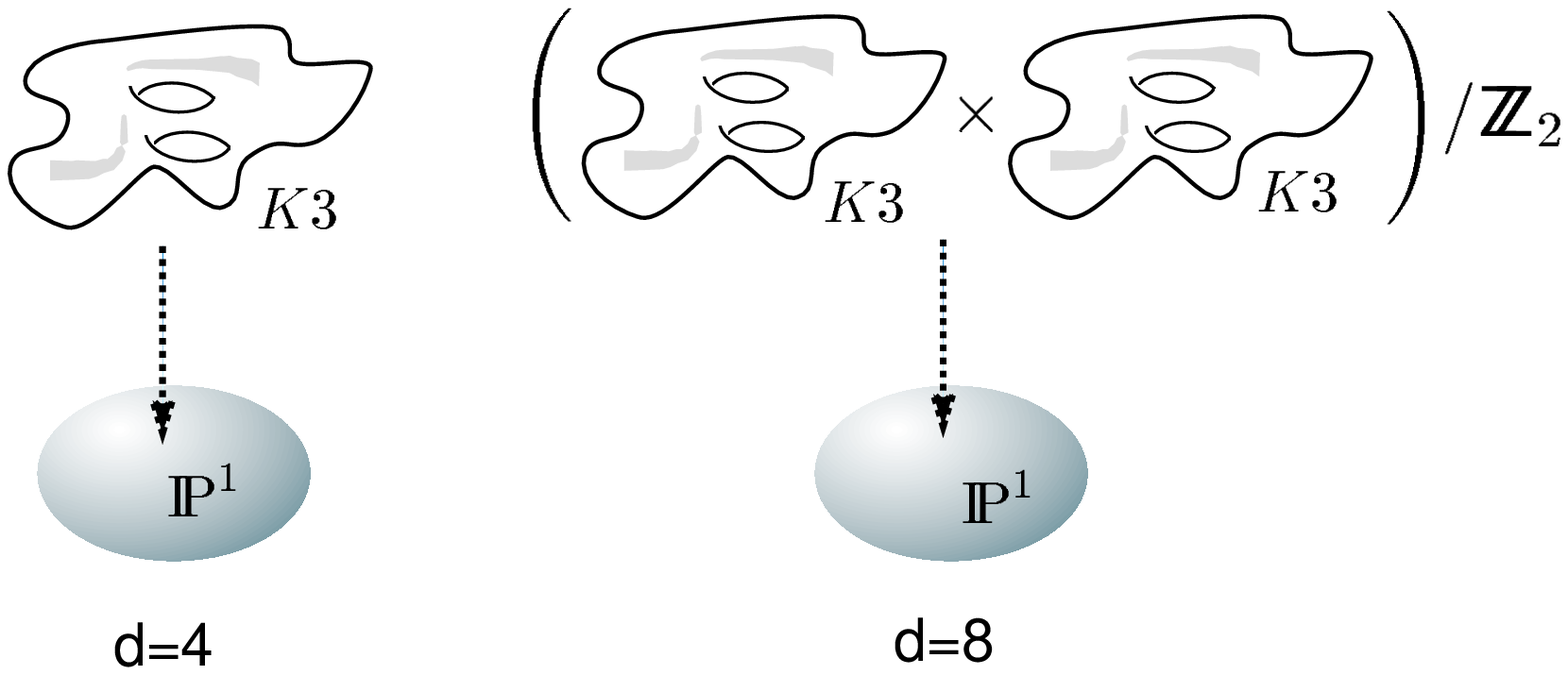}

The occurrence of such symmetric products is familiar in $D$-brane
physics. The geometrical structure  that is relevant to us is however
not just the symmetric square of $K3$, but rather a fibration of it, in
the limit of large base space -- this is precisely what the content of
the inhomogenous PF equation \eq{d=8inhom}\ is.  It is however not at
all obvious to us why this particular structure of a
hyperk\"ahler-fibered \index{five-fold}five-fold would underlie the
non-perturbative quantum geometry of the quartic gauge couplings in
eight dimensions.\footnote{The same structure is found also when
considering models with more moduli; in particular, the prepotential
$\cG(T,U)$ in \eq{prepot}\ can be re-derived in an analogous fashion
from symmetric squares \cite{LSWB}.}

The situation is, in this respect, somewhat similar to $N=2$ SYM theory
in four dimensions, where the Riemann surfaces underlying the effective
lagrangian were found in \cite{SW}, and at the time the geometry
appeared to be merely a convenient mathematical tool for encoding
appropriate physical data.  It was only quite some time later when  the
geometry was given a deep physical interpretation.\footnote{For
example, as part of world-volumina of type IIA \cite{KLMVW}\ or
$M$-theory \cite{WitM}\ fivebranes.} In the same spirit, one may
speculate that the five-folds that seem to emerge here may ultimately
have an interpretation in terms of a yet unknown dual formulation of
the theory, or, perhaps more likely, in terms of sigma-models
describing the relevant 7-brane interactions that lead to the requisite
$F^4$ terms in the effective action.

\section{Conclusion}

Summarizing, we have seen that the postulated duality between the
heterotic string compactified on $T^2$ and $F$-theory on $K3$ (which is
defined to be type IIB strings on $\IP^1$ with 24 $(p,q)$ 7-branes)
passes numerous tests. In particular, highly non-trivial quartic gauge
threshold coupling functions, which are one-loop exact on the heterotic
side, can be reproduced from geometrical data pertaining to $K3$
surfaces. Still, a better physical understanding of the issues
discussed in the previous paragraphs would be highly desirable.


\goodbreak
\begin{acknowledgments}
I thank Stephan Stieberger and Nick Warner for the
collaboration on this subject, and moreover the organizers of the
School for a very pleasant stay.
\end{acknowledgments}



\newcommand\nil[1]{{}}
\newcommand\nihil[1]{{\sl #1}}
\newcommand\ex[1]{}
\def\eprt#1{{\tt #1}}
\newcommand\br{{\hfill\break}}
\newcommand{\nup}[3]{{\em Nucl.\ Phys.}\ {B#1#2#3}\ }
\newcommand{\np}[3]{{\em Nucl.\ Phys.}\ {B#1#2#3}\ }
\newcommand{\plt}[3]{{\em Phys.\ Lett.}\ {B#1#2#3}\ }
\newcommand{\prl}{{\em Phys.\ Rev.\ Lett.}\ }
\newcommand{\cmp}[3]{{\em Comm.\ Math.\ Phys.}\ {A#1#2#3}\ }
\hfuzz=20pt



\lref\Kod{K.\ Kodaira, Ann.\ Math.\ {\bf 77} (1963) 563;
Ann.\ Math.\ {\bf 78} (1963) 1.}

\lref\alice{A.\ S.\ Schwarz,
\nihil{Field theories with no local 
conservation of the electric
charge,}
 Nucl.~ Phys.~{\bf B208} (1982) 141.}

\lref\ellg {A.\ Schellekens and N.\ Warner,
{\nihil{Anomalies, characters and strings,}
 Nucl.\  Phys.\ {\bf B287} (1987) 317;}\br
{E.\ Witten,
 \nihil{Elliptic genera and quantum field theory,}
 Comm.\  Math.\  Phys.\ {\bf 109} (1987) 525;}\br
W. Lerche, B.E.W. Nilsson, A.N. Schellekens and N.P. Warner,
 Nucl.\  Phys.\  {\bf 299} (1988) 91.}

\lref\DKLII{L. Dixon, V. Kaplunovsky and J. Louis,
 \nihil{Moduli dependence of string loop corrections to gauge coupling
constants,} Nucl.\ Phys.\ {\bf B355} (1991) 649-688.}

\lref\Fth{C.\ Vafa, 
\nihil{Evidence for F-Theory,}
 Nucl.\ Phys.\ {\bf B469} 403 (1996), 
\eprt{hep-th/9602022}. 
}

\lref\HM{J.A.\ Harvey and G.\ Moore, 
\nihil{Algebras, BPS States, and Strings,}
 Nucl.\ Phys.\ {\bf B463} 315 (1996), 
\eprt{hep-th/9510182}. 
}

\lref\WL{W. Lerche, {
 \nihil{Elliptic index and superstring effective actions,}
 Nucl.\  Phys.\ {\bf B308} (1988) 102.}}

\lref\Sen{A.\ Sen, 
\nihil{F-theory and Orientifolds,}
 Nucl.\ Phys.\ {\bf B475} 562 (1996), 
\eprt{hep-th/9605150}. 
}

\lref\BK{C.\ Bachas and E.\ Kiritsis, 
\nihil{$F^4$ terms in N = 4 string vacua,}
 Nucl.\ Phys.\ Proc.\ Suppl.\ {\bf 55B} 194 (1997), 
\eprt{hep-th/9611205}. 
}

\lref\BFKOV{C.\ Bachas, C.\ Fabre, E.\ Kiritsis, 
N.A.\ Obers and P.\ Vanhove, 
\nihil{Heterotic/type-I duality and D-brane instantons,}
 Nucl.\ Phys.\ {\bf B509} 33 (1998), 
\eprt{hep-th/9707126}. 
}

\lref\typeI{
{C.\ Bachas, 
\nihil{Heterotic versus type I,}
 Nucl.\ Phys.\ Proc.\ Suppl.\ {\bf 68} 348 (1998), 
\eprt{hep-th/9710102}; 
}
\br
{M.\ Bianchi, E.\ Gava, F.\ Morales and K.S.\ Narain, 
\nihil{D-strings in unconventional type I vacuum configurations,}
 Nucl.\ Phys.\ {\bf B547} 96 (1999), 
\eprt{hep-th/9811013}; 
}
\br
{E.\ Gava, A.\ Hammou, J.F.\ Morales and K.S.\ Narain, 
\nihil{On the perturbative corrections around D-string instantons,}
 JHEP{\bf 03} 023 (1999), 
\eprt{hep-th/9902202}; 
}
\br
{K.\ Foerger and S.\ Stieberger, 
\nihil{Higher derivative couplings and 
heterotic-type I duality in eight dimensions,}
\eprt{hep-th/9901020}; 
}
\br
{M.\ Gutperle, 
\nihil{A note on heterotic/type I
 duality and D0 brane quantum mechanics,}
 JHEP{\bf 05} 007 (1999), 
\eprt{hep-th/9903010}; 
}
\br
{E.\ Gava, K.S.\ Narain and M.H.\ Sarmadi, 
\nihil{Instantons in N = 2 Sp(N) superconformal 
gauge theories and the AdS/CFT correspondence,}
\eprt{hep-th/9908125}. 
}
}

\lref\OHirz{
{B.\ Craps and F.\ Roose, 
\nihil{Anomalous D-brane and orientifold couplings from the boundary state,}
 Phys.\ Lett.\ {\bf B445} 150 (1998), 
\eprt{hep-th/9808074}; 
}
\br
{B.J.\ Stefanski, 
\nihil{Gravitational couplings of D-branes and O-planes,}
 Nucl.\ Phys.\ {\bf B548} 275 (1999), 
\eprt{hep-th/9812088}. 
}
}

\lref\MSM{J.F.\ Morales, C.A.\ Scrucca and M.\ Serone, 
\nihil{Anomalous couplings for D-branes and O-planes,}
 Nucl.\ Phys.\ {\bf B552} 291 (1999), 
\eprt{hep-th/9812071}. 
}

\lref\GHM{M.\ Green, J.A.\ Harvey and G.\ Moore, 
\nihil{I-brane inflow and anomalous couplings on D-branes,}
 Class.\ Quant.\ Grav.\ {\bf 14} 47 (1997), 
\eprt{hep-th/9605033}. 
}

\lref\KDSM{K.\ Dasgupta and S.\ Mukhi, 
\nihil{F-theory at constant coupling,}
 Phys.\ Lett.\ {\bf B385} 125 (1996), 
\eprt{hep-th/9606044}. 
}

\lref\DJM{K.\ Dasgupta, D.P.\ Jatkar and S.\ Mukhi, 
\nihil{Gravitational couplings and Z(2) orientifolds,}
 Nucl.\ Phys.\ {\bf B523} 465 (1998), 
\eprt{hep-th/9707224}. 
}

\lref\ko{E.\ Kiritsis and N.A.\ Obers, 
\nihil{Heterotic type I duality in $d <10$-dimensions, threshold corrections and D-instantons,}
JHEP{\bf 10} 004 (1997), 
\eprt{hep-th/9709058}. 
}

\lref\CV{S.\ Cecotti and C.\ Vafa, 
\nihil{Topological-antitopological fusion,}
 Nucl.\ Phys.\ {\bf B367} 359 (1991). 
}

\lref\junct{See eg.:\br
{A.\ Johansen, 
\nihil{A comment on BPS states in F-theory in 8 dimensions,}
 Phys.\ Lett.\ {\bf B395} 36 (1997), 
\eprt{hep-th/9608186}; 
}
\br
{M.R.\ Gaberdiel and B.\ Zwiebach, 
\nihil{Exceptional groups from open strings,}
 Nucl.\ Phys.\ {\bf B518} 151 (1998), 
\eprt{hep-th/9709013}. 
}
\br
{M.R.\ Gaberdiel, T.\ Hauer and B.\ Zwiebach, 
\nihil{Open string-string junction transitions,}
 Nucl.\ Phys.\ {\bf B525} 117 (1998), 
\eprt{hep-th/9801205}; 
}
\br
{O.\ DeWolfe and B.\ Zwiebach, 
\nihil{String junctions for arbitrary Lie algebra representations,}
 Nucl.\ Phys.\ {\bf B541} 509 (1999), 
\eprt{hep-th/9804210}. 
}
}

\lref\WLSS{W.\ Lerche and S.\ Stieberger, 
\nihil{Prepotential, mirror map and F-theory on K3,}
 Adv.\ Theor.\ Math.\ Phys.\ {\bf 2} 1105 (1998), 
\eprt{hep-th/9804176}. 
}

\lref\LSWA{W.\ Lerche, S.\ Stieberger and N.P.\ Warner, 
\nihil{Quartic gauge couplings from K3 geometry,}
\eprt{hep-th/9811228}. 
}

\lref\LSWB{W.\ Lerche, S.\ Stieberger and N.P.\ Warner, 
\nihil{Prepotentials from symmetric products,}
\eprt{hep-th/9901162}. 
}

\lref\CB{C.\ Bachas, 
\nihil{(Half) a lecture on D-branes,}
\eprt{hep-th/9701019}. 
}

\lref\dieter{G.\ Lopes Cardoso, D.\ Lust and T.\ Mohaupt, 
\nihil{Threshold corrections and symmetry 
enhancement in string compactifications,}
 Nucl.\ Phys.\ {\bf B450} 115 (1995), 
\eprt{hep-th/9412209}. 
}

\lref\enhancements{
{B.\ de Wit, V.\ Kaplunovsky, J.\ Louis and D.\ Lust, 
\nihil{Perturbative couplings of vector 
multiplets in N=2 heterotic string vacua,}
 Nucl.\ Phys.\ {\bf B451} 53 (1995), 
\eprt{hep-th/9504006}; 
}\br
{I.\ Antoniadis, S.\ Ferrara, E.\ Gava, K.S.\ Narain and T.R.\ Taylor, 
\nihil{Perturbative prepotential and monodromies 
in N=2 heterotic superstring,}
 Nucl.\ Phys.\ {\bf B447} 35 (1995), 
\eprt{hep-th/9504034}. 
}
}

\lref\GSVY{B.R.\ Greene, A.\ Shapere, C.\ Vafa and S.\ Yau, 
\nihil{Stringy Cosmic Strings And Noncompact Calabi-Yau Manifolds,}
 Nucl.\ Phys.\ {\bf B337} 1 (1990). 
}

\lref\LSW{W.\ Lerche, D.J.\ Smit and N.P.\ Warner, 
\nihil{Differential equations for periods and flat coordinates in two-dimensional topological matter theories,}
 Nucl.\ Phys.\ {\bf B372} 87 (1992), 
\eprt{hep-th/9108013}. 
}

\lref\LY{
{B.H.\ Lian and S.\ Yau, 
\nihil{Arithmetic properties of mirror map and quantum coupling,}
 Commun.\ Math.\ Phys.\ {\bf 176} 163 (1996), 
\eprt{hep-th/9411234}, 
}
{ 
\nihil{Mirror maps, modular relations and hypergeometric series I,}
\eprt{hep-th/9507151}, 
}
{ 
\nihil{Mirror maps, modular relations and hypergeometric series.\ II,}
 Nucl.\ Phys.\ Proc.\ Suppl.\ {\bf 46} 248 (1996), 
\eprt{hep-th/9507153}. 
}
}

\lref\DVV{R.\ Dijkgraaf, H.\ Verlinde and E.\ Verlinde, 
\nihil{Topological strings in $d<1$},
Nucl.\ Phys.\ {\bf B352} 59 (1991). 
}

\lref\Rankin{R.\ Rankin, \nihil{Modular forms and functions},
Cambridge University Press.}

\lref\ZN{
{L.\ Dixon, D.\ Friedan, E.\ Martinec and S.\ Shenker,
 \nihil{The conformal field theory of orbifolds,}
 Nucl.\  Phys.\ {\bf B282} (1987) 13;}
\br
{M.\ Bershadsky and A.\ Radul,
\nihil{Conformal field theories with additional Z(N) symmetry,} 
Sov.\ J.\ Nucl.\ Phys.\ 47 (1988) 363-369;}
\br
{F.\ Ferrari and J.\ Sobczyk, 
\nihil{Bosonic Field Propagators on Algebraic Curves,}
\eprt{hep-th/9909173}. 
}
}

\lref\barrozo{M.C.\ Barrozo, 
\nihil{Map of Heterotic and Type IIB Moduli in 8 Dimensions,}
\eprt{hep-th/9909178}. 
}

\lref\SW{N.\ Seiberg and E.\ Witten, 
\nihil{Electric - magnetic duality, monopole condensation, 
and confinement in N=2 supersymmetric Yang-Mills theory,}
 Nucl.\ Phys.\ {\bf B426} 19 (1994), 
\eprt{hep-th/9407087}. 
}

\lref\KV{S.\ Kachru and C.\ Vafa, 
\nihil{Exact results for N=2 compactifications of heterotic strings,}
 Nucl.\ Phys.\ {\bf B450} 69 (1995), 
\eprt{hep-th/9505105}. 
}

\lref\KLM{A.\ Klemm, W.\ Lerche and P.\ Mayr, 
\nihil{K3 Fibrations and heterotic type II string duality,}
 Phys.\ Lett.\ {\bf B357} 313 (1995), 
\eprt{hep-th/9506112}. 
}

\lref\KKLMV{S.\ Kachru, A.\ Klemm, W.\ Lerche, P.\ Mayr and C.\ Vafa, 
\nihil{Nonperturbative results on the point particle limit of 
N=2 heterotic string compactifications,}
 Nucl.\ Phys.\ {\bf B459} 537 (1996), 
\eprt{hep-th/9508155}. 
}

\lref\PADMKthree{P.S.\ Aspinwall and D.R.\ Morrison, 
\nihil{String theory on K3 surfaces,}
\eprt{hep-th/9404151}. 
}

\lref\AL{P.S.\ Aspinwall and J.\ Louis, 
\nihil{On the Ubiquity of K3 Fibrations in String Duality,}
 Phys.\ Lett.\ {\bf B369} 233 (1996), 
\eprt{hep-th/9510234}. 
}

\lref\PAK3{P.S.\ Aspinwall, 
\nihil{K3 surfaces and string duality,}
\eprt{hep-th/9611137}. 
}

\lref\mirror{
See e.g.,
\nihil{Essays and mirror manifolds}, (S.\ Yau, ed.),
International Press 1992;
\nihil{Mirror symmetry II}, (B.\ Greene et al, eds.),
International Press 1997.
}

\lref\symm2{
{M.\ Lee,
 \nihil{Picard-Fuchs equations for elliptic modular varieties,}
 Appl.\ Math.\ Letters 4, no.5 (1991) 91-95;
}\br
{C.\ Doran,
 \nihil{Picard-Fuchs Uniformization:
  Modularity of the Mirror Map and Mirror-Moonshine},
 \eprt{math.AG/9812162}.}
}

\lref\KLMVW{A.\ Klemm, W.\ Lerche, P.\ Mayr, C.\ Vafa and N.\ Warner, 
\nihil{Self-Dual Strings and N=2 Supersymmetric Field Theory,}
 Nucl.\ Phys.\ {\bf B477} 746 (1996), 
\eprt{hep-th/9604034}. 
}

\lref\HKTY{S.\ Hosono, A.\ Klemm, S.\ Theisen and S.T.\ Yau, 
\nihil{Mirror symmetry, mirror map and applications to Calabi-Yau 
hypersurfaces,}
 Commun.\ Math.\ Phys.\ {\bf 167} 301 (1995), 
\eprt{hep-th/9308122}. 
}

\lref\WitM{E.\ Witten, 
\nihil{Solutions of four-dimensional field theories via M-theory,}
 Nucl.\ Phys.\ {\bf B500} 3 (1997), 
\eprt{hep-th/9703166}. 
}

\lref\hyperk{See e.g.,
{D.\ Huybrechts,
\nihil{Compact hyperk\"ahler manifolds: basic results},
\eprt{alg-geom/9705025}; }\br
{R.\ Dijkgraaf,
 \nihil{Instanton strings and hyperK\"ahler geometry,}
 \eprt{hep-th/9810210}.}
}

\lref\CCLM{G.\ Lopes Cardoso, G.\ Curio, D.\ L\"ust and T.\ Mohaupt, 
\nihil{On the duality between the heterotic 
string and F-theory in 8 dimensions,}
 Phys.\ Lett.\ {\bf B389} 479 (1996), 
\eprt{hep-th/9609111}. 
}

\lref\JS{J.H.\ Schwarz, 
\nihil{An SL(2,Z) multiplet of type IIB superstrings,}
 Phys.\ Lett.\ {\bf B360} 13 (1995), 
\eprt{hep-th/9508143}. 
}

\lref\WLSSKod{W.\ Lerche and S.\ Stieberger, 
\nihil{On the anomalous and global interactions of Kodaira 7-planes,}
\eprt{hep-th/9903232}. 
}



\end{document}